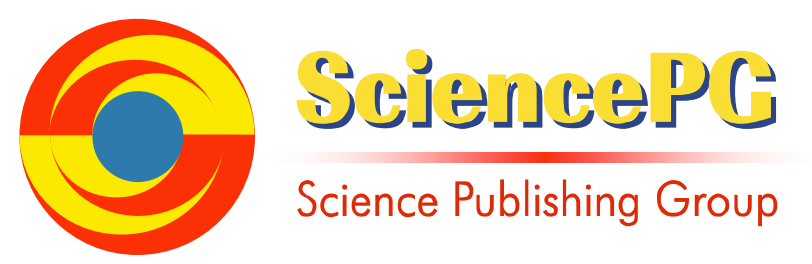

**Research/Technical Note**

# Radiation Analysis for Moon and Mars Missions

**Andreas Märki**

Märki Analytics for Space, Erlenbach ZH, Switzerland

**Email address:**
a.m.maerki@bluewin.ch

**Abstract:** This paper provides an overview of the radiation aspects of manned space flight to Moon and Mars. The expected ionizing radiation dose for an astronaut is assessed along the Apollo 11 flight path to the Moon. With the two dose values, the expected and the measured total dose, the radiation shielding and the activity of the Sun are estimated. To judge the risk or safety margin the radiation effects on humans are opposed. The radiation from the Sun has to be set to zero in the computer model to achieve the published radiation dose value of the Apollo 11 flight. Galactic and cosmic particles have not been modelled either. The Apollo 11 astronauts must have been lucky that during their flight the Sun was totally quiet in the solar maximum year 1969 – and also their colleagues of the subsequent Apollo flights, i.e. until 1972, where the published dose values still require a quiet Sun. The here built mathematical model allows assessing the total dose of a journey to Mars by only changing the flight duration. Even if in the meantime much thicker and/or active radiation shielding is proposed the radiation risk of manned space flight to Moon and Mars remains still huge.

**Keywords:** Space Radiation, Van Allen Belt, Apollo 11

## 1. Introduction

Electronics which is used in space has to be specifically robust so that it sustains the radiation environment. Often it is packed in a metallic housing to achieve the necessary radiation shielding. Humans are much more susceptible to ionizing radiation than electronics.

The flight path to the Moon or to Mars crosses the Van Allen radiation belt (VAB, Figure 1), a zone with free protons and electrons. These charged particles produce a strong radiation so that space crafts need an appropriate shielding. After this belt the space craft is directly exposed to the solar radiation because there is no protection of the Earth magnetic field any more. Nevertheless the absorbed radiation dose of an Apollo 11 astronaut was only 0.18 rad or 1.8 mGy (Milli-Gray). This is the minimum dose of the Apollo flights; the maximum is 1.14 rad (Apollo 14) [1].

After Apollo NASA stated that "*radiation was not an operational problem during the Apollo Program*." [13]

But in 1997 NASA proposed to massively increase the radiation shielding for lunar missions [19] and ESA started to study active radiation protection in 2011 [20]. An article published in 2018 shows the still poor knowledge about the radiation risk for travelling beyond the Low Earth Orbits (LEO) [21].

The goal of this analysis is to evaluate under which conditions the Apollo dose values could occur and whether these dose values can be used for the estimation of the radiation risk for future missions beyond LEO. Further an estimation of the radiation dose of future Moon and Mars missions shall be made.

## 2. Radiation Data

The 0.18 rad of Apollo 11 correspond to 1.8 mGy or (optimistically) to 1.8 mSv (Milli-Sievert). Sievert indicates biological effects; depending on the kind of radiation and tissue there is a weighting factor of > 1 to be considered for the conversion from Gray to Sievert. If Grays are 1:1 converted to Sievert then in general there results a too low dose value in Sievert. Here I use the 1:1conversion.

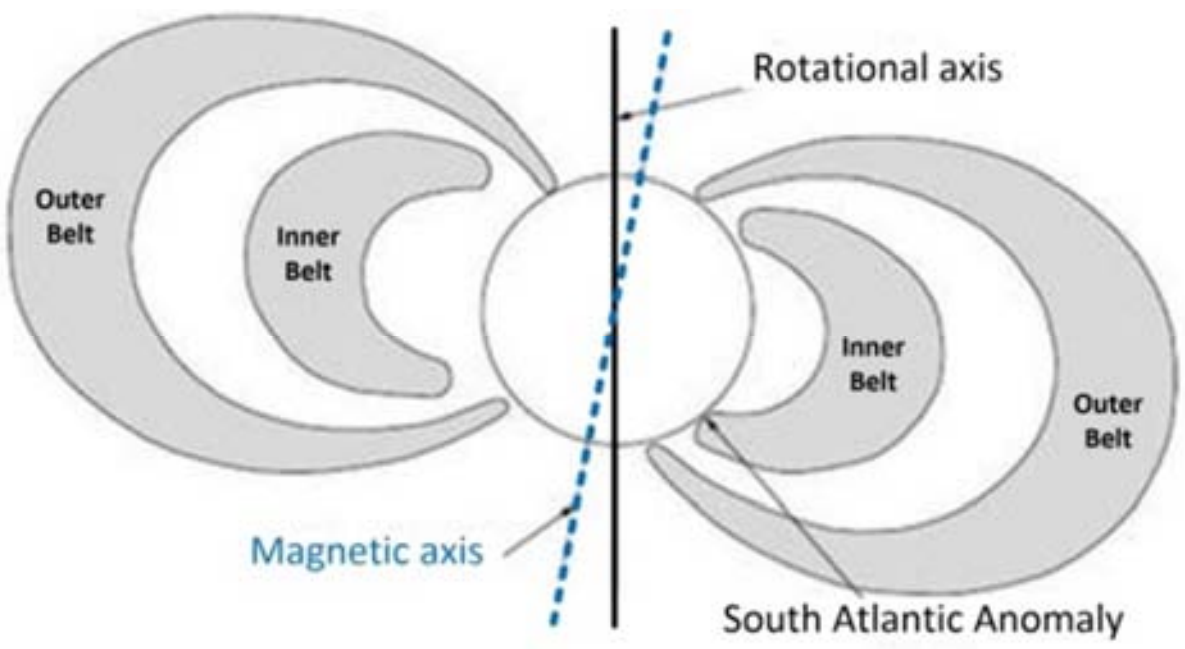

***Figure 1.*** *Van Allen Radiation Belt [6].*

As a comparison the average annual terrestrial dose level is according to [2] 2.4 mSv, other sources show 2.5 - 4.5 mSv/year. This means, the 1.8 mSv are a small additional exposure.

Another example would be an astronaut in a space craft orbiting the Earth 1'000 km above the equator. If the space craft were protected with 4 mm aluminium shielding then he would be exposed to a dose rate of 2.7 mSv/h [3]. This would be about one natural annual dose per hour.

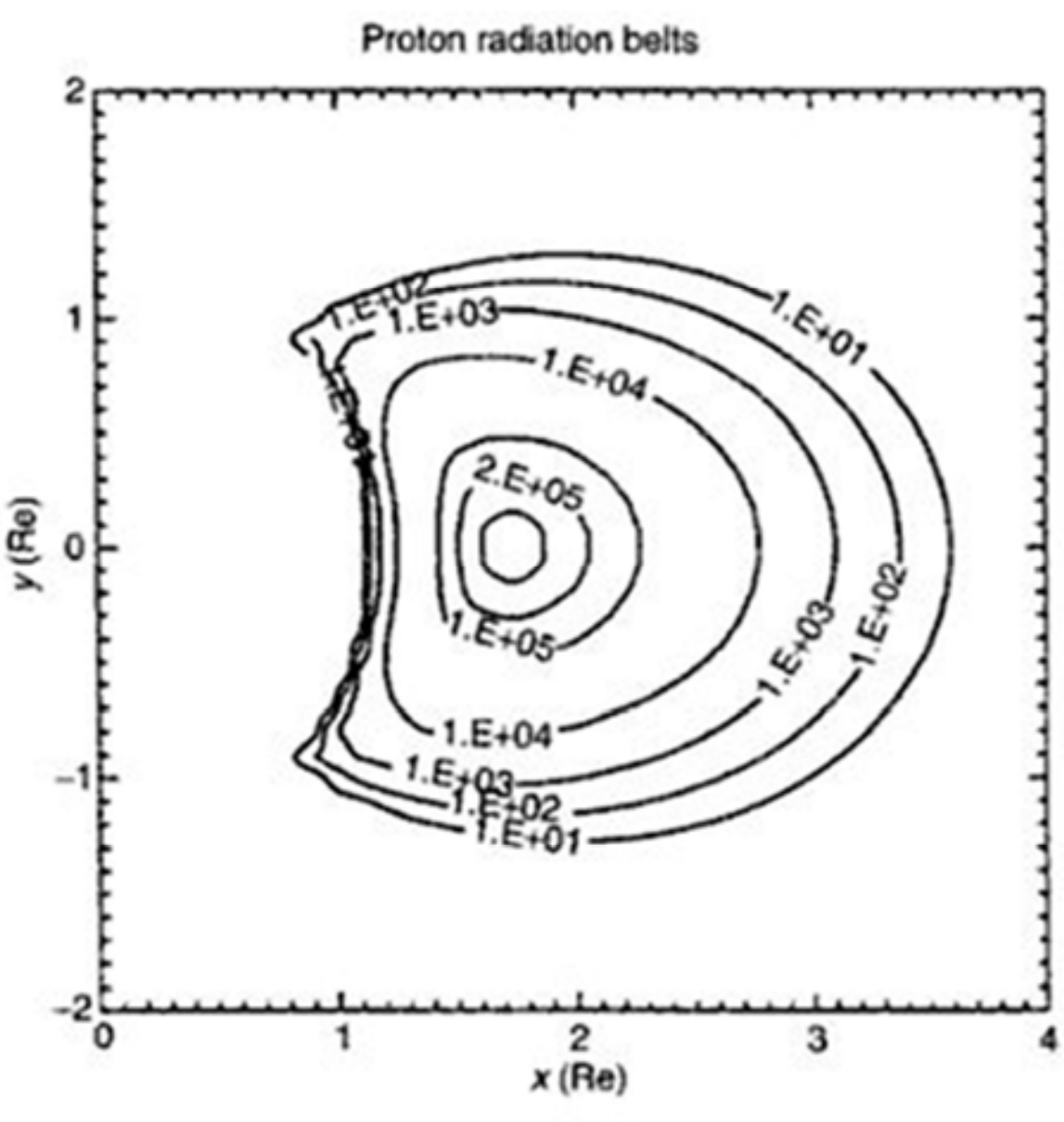

Re means Radius of the Earth (6370km)

***Figure 2.*** *Flux (number/(cm$^2$·s)) of high Energy Protons [4].*

The explanation for this high level is the Van Allen radiation belt which encircles the Earth: at a few hundred kilometres altitude the radiation rapidly grows. At 1'000 km above the equator it is – as shown before – quite high and it increases further. At 3'000 km above the equator the dose rate is 465 mSv/h (always with 4 mm aluminium shielding), and only after 40'000 km the dose rate falls below the value of 1'000 km altitude.

To fly to the Moon and back or to Mars one does not have to cross the Van Allen radiation belt through its centre, but it generally takes more than one hour to cross it.

Figures 2 to 5 show the situation in the Van Allen radiation belt more in detail: Figures 2 and 3 are diagrams with level curves, which indicate the number of high energy particles, and Figures 4 and 5 show the total annual dose as a function of the altitude above the equator.

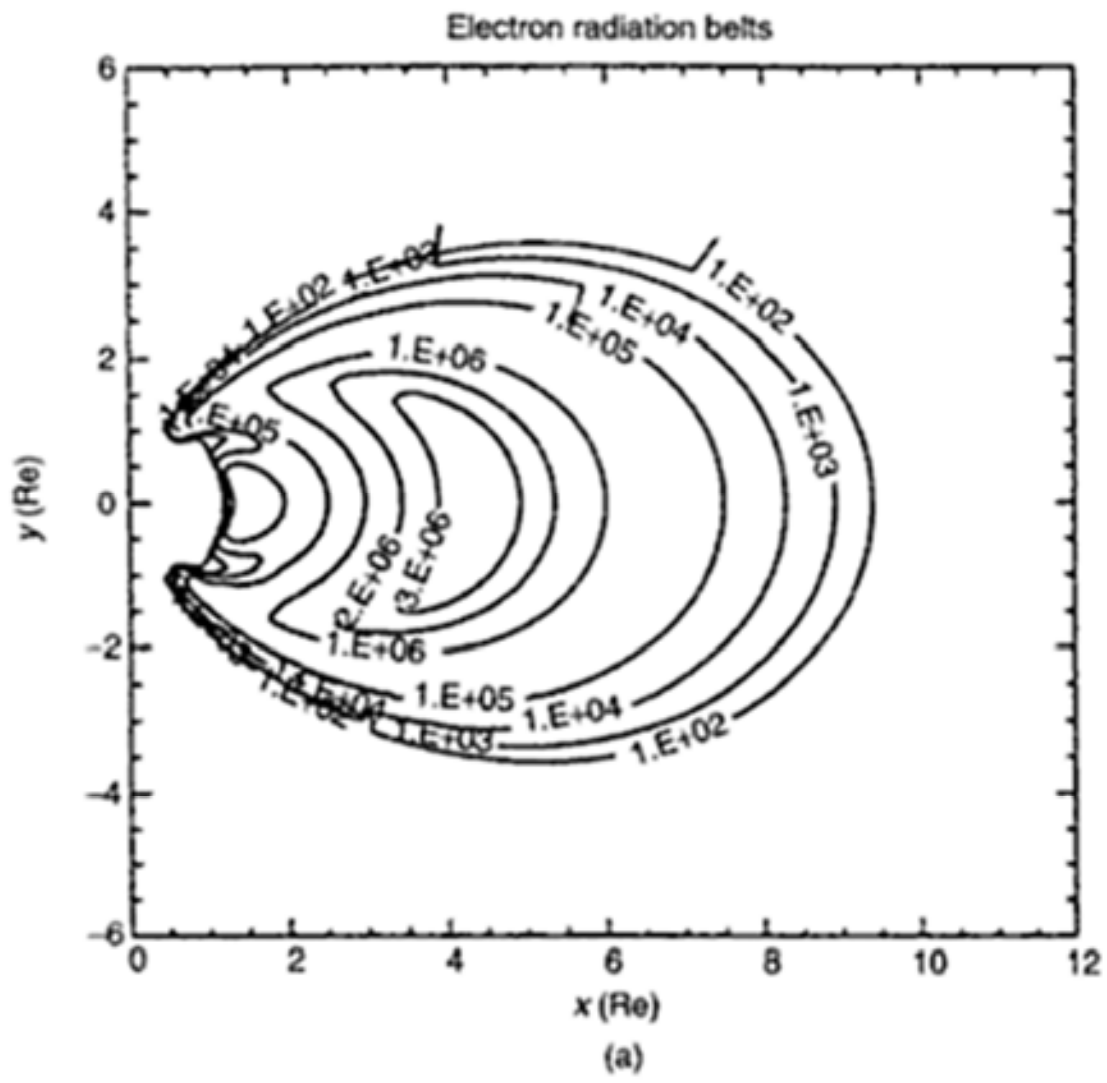

***Figure 3.*** *Flux (number/(cm$^2$·s)) of high Energy Electrons [4].*

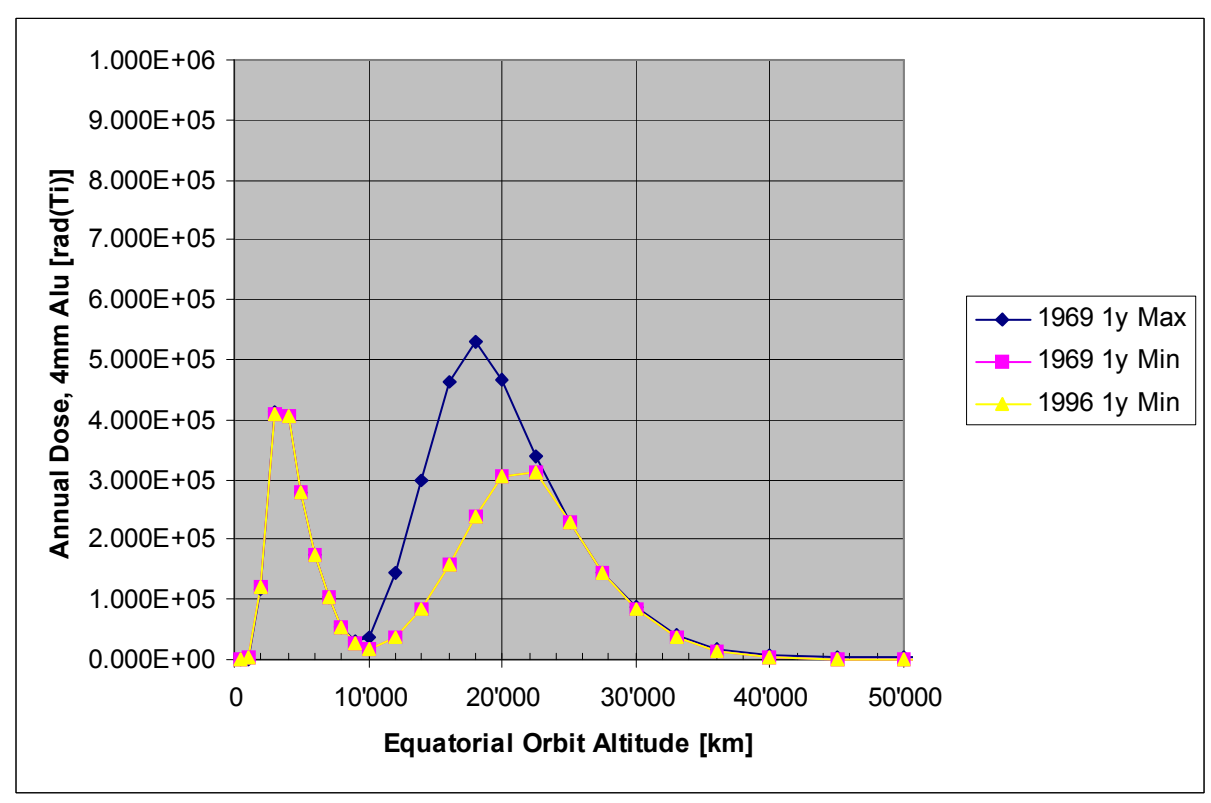

***Figure 4.*** *Annual Dose in the Van Allen Radiation Belt; determined with [3]. Linear scaling; identical data as in Figure 5.*

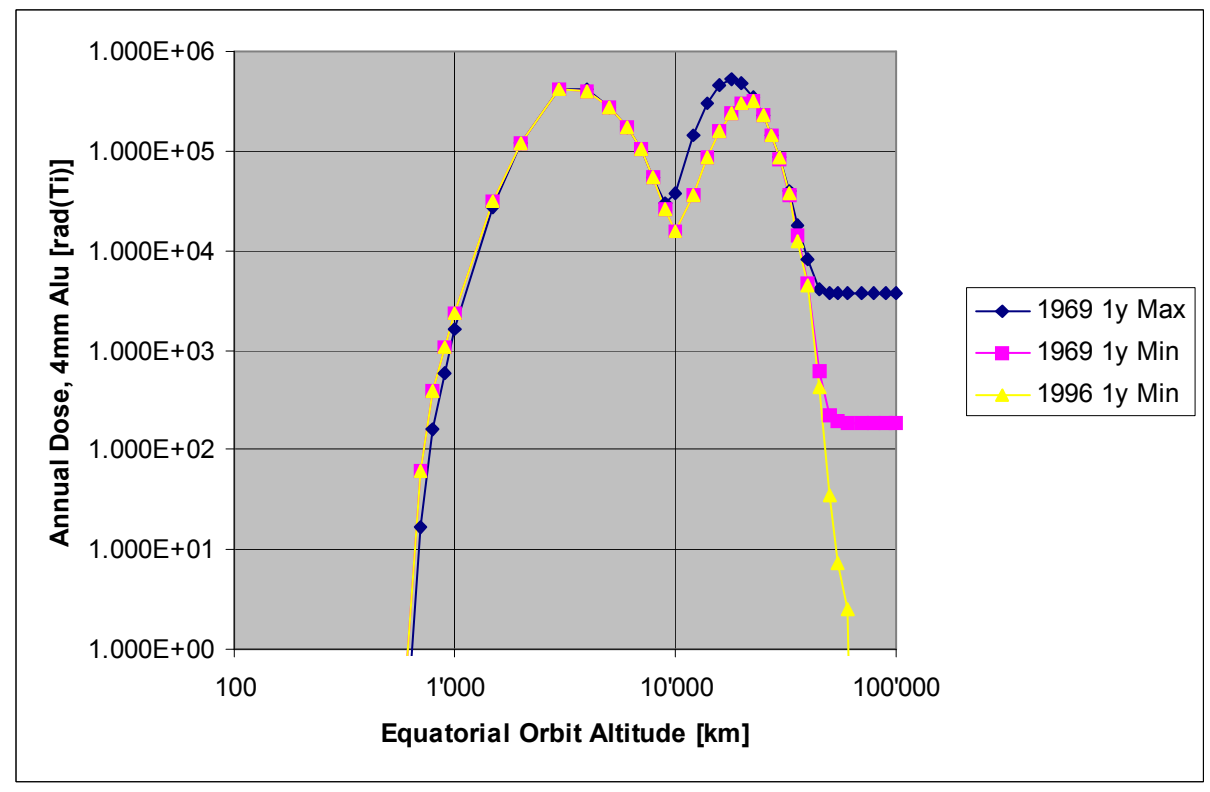

***Figure 5.*** *Annual Dose in the Van Allen Radiation Belt; determined with [3]. The horizontal lines on the right are contributions from the Sun.*

The diagram of Figure 5 corresponds well with similar diagrams in [4, 14]. The dose level is shown for tissue (Ti), on the comparative diagrams it is shown for silicon (Si). For tissue the dose level is about 30% higher. But this is hardly visible on the logarithmic scale.

The dose levels of Figure 4 and 5 refer to equatorial orbits. The equatorial plane is inclined with respect to the symmetry plane of the Van Allen radiation belt (i.e. the plane perpendicular to the magnetic axis, see Figure 1 or Figure 6), so that equatorial orbits are not always in the area of the maximum radiation. This means that in the centre of the Van Allen radiation belt there are even higher dose rates than shown in the above Figures 4 and 5.

With the annual doses one can determine the total dose of a mission only then exactly if the radiation over the year is constant. In the Van Allen radiation belt there is a constant (base) radiation of the free electrons and protons which are always there. Outside of the belt there is variable proton radiation from the Sun. Further radiation is neglected here.

On the way to the Moon, which is investigated first, the ionizing radiation depends mainly on the activity of the Sun. If the Sun is not active its radiation is almost zero. At an eruption or flare up to 10 Sv can occur per event [11], what *"in fact should be deadly."* [21] Mentions peak dose rates of 2'837 mGy/h, i.e. the same order of magnitude.

Here a constant radiation beyond the VAB up to the Moon is assumed; an astronaut flying to the Moon is exposed to this radiation over several days.

Averaged over the year 1969 the total radiation dose beyond the Van Allen radiation belt was significant, as it is shown in Table 1.

These dose values are determined from the solar parts of Figure 5, "1969 1y Max and Min". At 4 mm Alu shielding the upper part adds up to about 3700 rad/y, which corresponds to ≥37 Sv/y or ≥4.2 mSv/h, the lower part adds up to about 190 rad/y or ≥1.9 Sv/y or ≥0.2 mSv/h.

The values from Table 1 exceed the dose of the Apollo 11 mission (1.8 mGy) by far – even for a 20 mm shielding. They show the serious risk, i.e. what could happen if the Sun were active as usual. By the way, even today a solar eruption cannot be predicted, not even one day in advance.

But these average values may yield wrong values for short missions. According to SPENVIS [3] the radiation dose from the Sun originated in the year 1969 mainly from a short active phase:

*"Cycle 20 had one anomalously large event that accounted for most of the accumulated fluence."* (Cycle 20: 1966-1972)

Because of the unique event in the solar cycle of 1966-1972 one could assume that during the Apollo 11 mission only little or even no (proton) radiation was radiated from the Sun. For this reason a totally radiation-inactive Sun is assumed in the following and only the base VAB radiation, which is always present, is considered, i.e. the yellow curve in Figures 4 and 5.

## 3. Estimation of the Radiation Dose according to an Idealised Straight Flight Path

In this chapter I describe a heuristic approach to estimate the dose level assuming a straight trajectory with an average speed. The trajectory is in the ecliptic, i.e. in the plane of the Earth orbit around the Sun. I will refine this approach in the next chapter.

To determine the radiation dose of the Apollo 11 mission, I always take the smallest value for the total dose per investigated area. So at the end the result is a lower limit of the total dose. A reason for this is the total dose value of 1.8 mGy which looks small. Therefore I check whether under favourable circumstances such a low level can be obtained.

The approach is as follows:

1. The Van Allen radiation belt shall be crossed at its border. I assume a maximum crossing angle of 35°:

a) The Earth axis is 23.5° inclined relative to the ecliptic.

b) The magnetic pole was in 1969 11.5° displaced relative to the geographical North Pole ([8], see also Figure 1).

Therefore the Van Allen radiation belt may have had its maximum inclination of 35° with respect to the ecliptic.

2. In the radiation determination program [3] all switches are now set to "Minimum" to achieve as small as possible total dose levels. In particular I tuned the program so that there is no contribution from the Sun at all. For this I selected the solar minimum year 1996 instead of 1969! "No contribution from the Sun" can be correct for a certain period of time, and in a year of a solar minimum this is the normal case. But for the solar maximum year 1969 [3] this assumption is very optimistic, even if in that year the solar activity was smaller compared to earlier solar maximum years and if the main part of the total dose in that cycle came from one single event – what, by the way, was not known in 1969.

Summarised only the part of the radiation in the Van Allen radiation belt is considered which is always present, i.e. the radiation of the "trapped" free protons and electrons.

3. The total dose of Figures 4 and 5 is calculated for equatorial orbits, it therefore corresponds to a mean value in a cone of ±11.5° (in Figures 7 and 8). In the centre of the Van Allen radiation belt the total dose value is higher. All the same I take the too low mean value for the central value and remain on the conservative side.

4. The radiation dose is initially calculated in rad or Gray. I make a 1:1 conversion from Gray to Sievert, i.e. there results a too low dose value in Sievert. For the comparison with the mission value this has no impact, because the mission value is given in rad (1 rad=10mGy). Sievert is used for biological effects.

***Table 1.*** *Expected Total Dose outside the VAB up to the Moon, averaged for 1969, determined with [3].*

| Shielding [mm Aluminium] | High particle fluences from the Sun ("Max" or confidence level=95%) | | Low particle fluences from the Sun ("Min" or confidence level=50%) | |
|---|---|---|---|---|
| | **Dose over 2h (on the Moon) [mSv]** | **(Travel) Dose over 180h (7.5d) [mSv]** | **Dose over 2h (on the Moon) [mSv]** | **(Travel) Dose over 180h (7.5d) [mSv]** |
| 0.05 | 810.18 | 72'854.2 | 217.57 | 19'564.7 |
| 0.1 | 415.70 | 37'392.2 | 77.80 | 6'995.9 |

| Shielding [mm Aluminium] | High particle fluences from the Sun (“Max” or confidence level=95%) | | Low particle fluences from the Sun (“Min” or confidence level=50%) | |
|---|---|---|---|---|
| | Dose over 2h (on the Moon) [mSv] | (Travel) Dose over 180h (7.5d) [mSv] | Dose over 2h (on the Moon) [mSv] | (Travel) Dose over 180h (7.5d) [mSv] |
| 0.2 | 215.86 | 19'410.7 | 30.07 | 2'704.3 |
| 0.3 | 150.10 | 13'496.9 | 18.21 | 1'638.0 |
| 0.4 | 113.83 | 10'236.1 (1 krad) | 12.51 | 1'125.3 (110 rad) |
| 0.5 | 90.46 | 8'135.5 | 9.16 | 823.6 |
| 0.6 | 74.29 | 6'681.7 | 7.00 | 629.0 |
| 0.8 | 54.64 | 4'913.8 | 4.63 | 416.4 |
| 1 | 43.24 | 3'889.1 | 3.41 | 306.4 |
| 1.5 | 27.93 | 2'509.2 | 1.92 | 172.8 |
| 2 | 20.14 | 1'810.9 | 1.27 | 114.3 |
| 2.5 | 15.42 | 1'386.9 | 0.91 | 81.7 |
| 3 | 12.20 | 1'097.3 (110 rad) | 0.68 | 60.8 |
| 4 | 8.43 | 757.7 | 0.43 | 38.3 |
| 5 | 6.24 | 561.6 | 0.30 | 26.5 |
| 6 | 4.96 | 446.2 | 0.22 | 20.0 |
| 7 | 4.01 | 360.8 | 0.17 | 15.5 |
| 8 | 3.34 | 300.4 | 0.14 | 12.5 |
| 9 | 2.86 | 257.5 | 0.12 | 10.4 |
| 10 | 2.45 | 220.7 | 0.10 | 8.7 |
| 12 | 1.91 | 171.5 | 0.07 | 6.5 |
| 14 | 1.51 | 135.5 | 0.05 | 4.9 |
| 16 | 1.23 | 110.6 | 0.04 | 3.9 |
| 18 | 1.03 | 92.9 | 0.04 | 3.2 |
| 20 | 0.87 | 77.9 | 0.03 | 2.6 |

Figure 6 shows the constellation of the magnetic axis with the maximum inclination of the Van Allen radiation belt with respect to the ecliptic. The ecliptic is shown in red. The green ellipse shows the short path through the VAB.

The angle between the blue solid line, which is perpendicular to the magnetic axis, and the red line, the ecliptic, is 35° as described above.

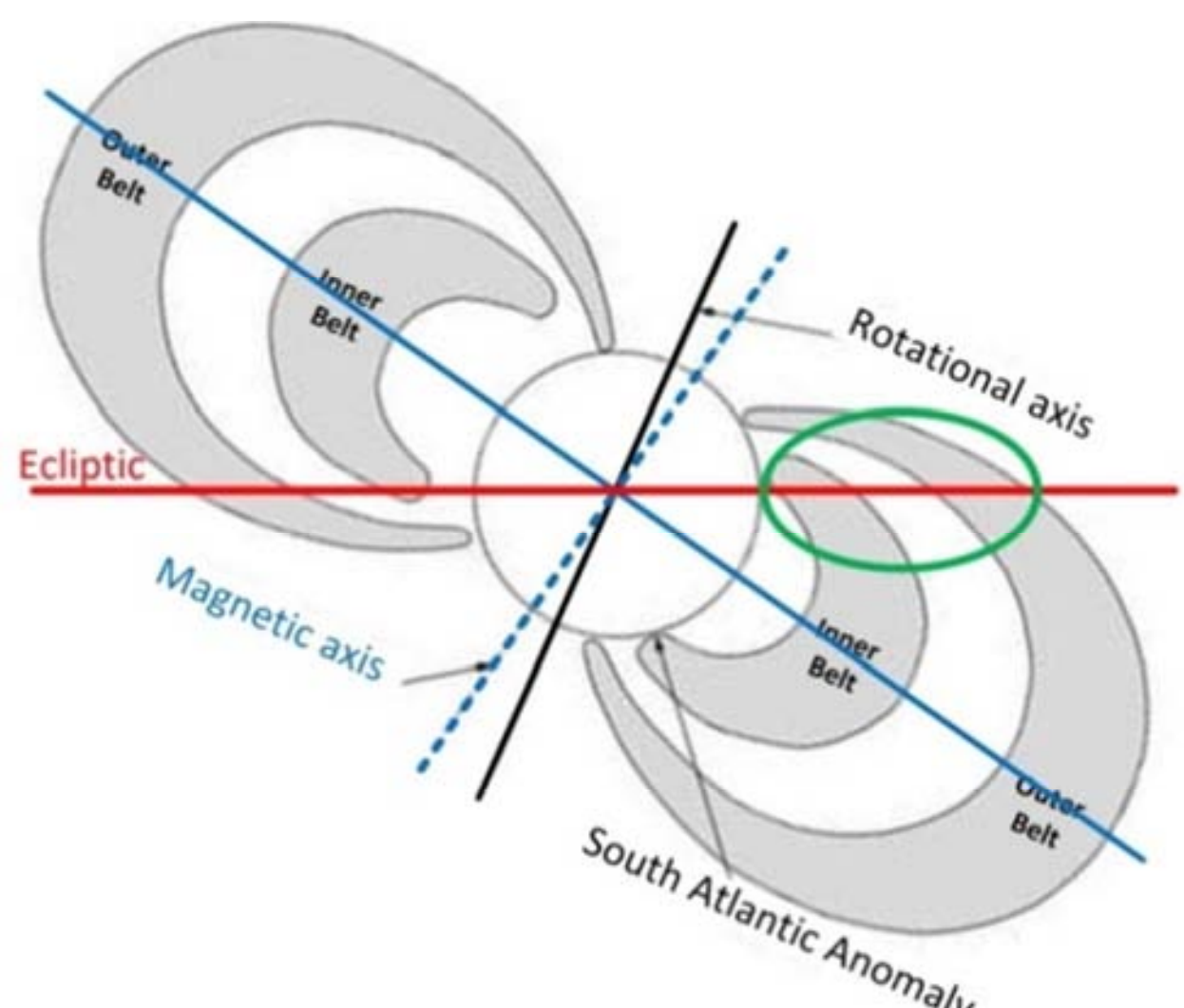

***Figure 6.*** *Van Allen Radiation Belt at its maximum Inclination w.r.t. the Ecliptic.*

The lunar orbit is inclined 5° with respect to the ecliptic, but the Moon was at the arrival of Apollo 11 quite exactly in the ecliptic [9] so that the flight path is also about in the ecliptic. These 5° may therefore not be added to further shorten the flight path through the VAB in the case of Apollo 11.

Apollo 11 could have penetrated the Van Allen radiation belt in the best case under an angle of about 35° to get a minimum radiation dose. Figure 7 and Figure 8 show the situation for protons and electrons separately.

The path length in the zone with more than 1E+05 high energy electrons (yellow marked) is 3.1·Re ≈ 20'000km.

40% of the yellow part are in the zone with more than 1E+06 high energy electrons/(cm$^2$·s).

The flight path is tangential to the proton belt and avoids the zone with the maximum radiation. In the electron belt it still avoids the centre, but it crosses a zone with rather high energy electrons.

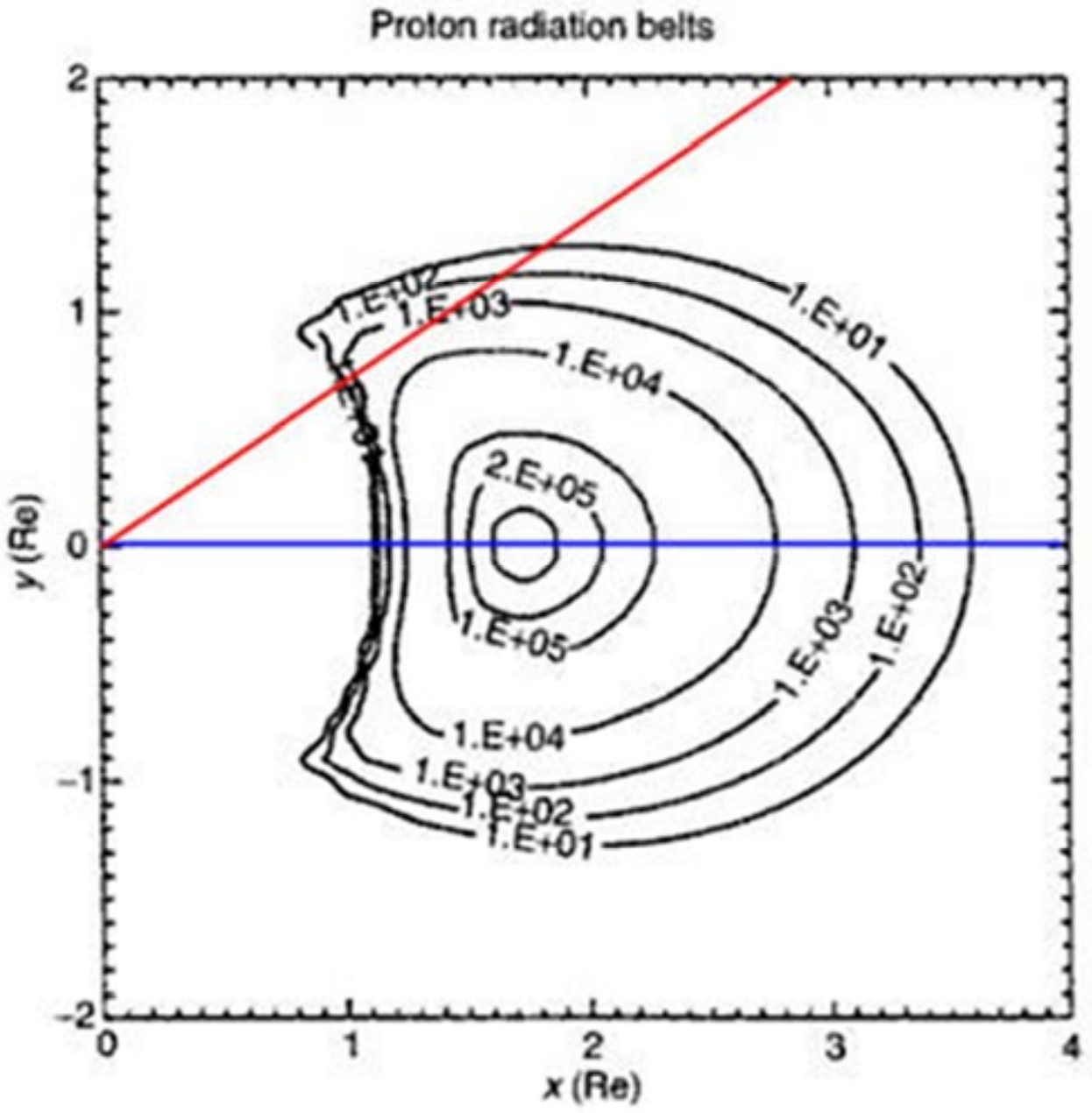

***Figure 7.*** *The Inner Part of the VAB: The Proton Radiation Belt.*

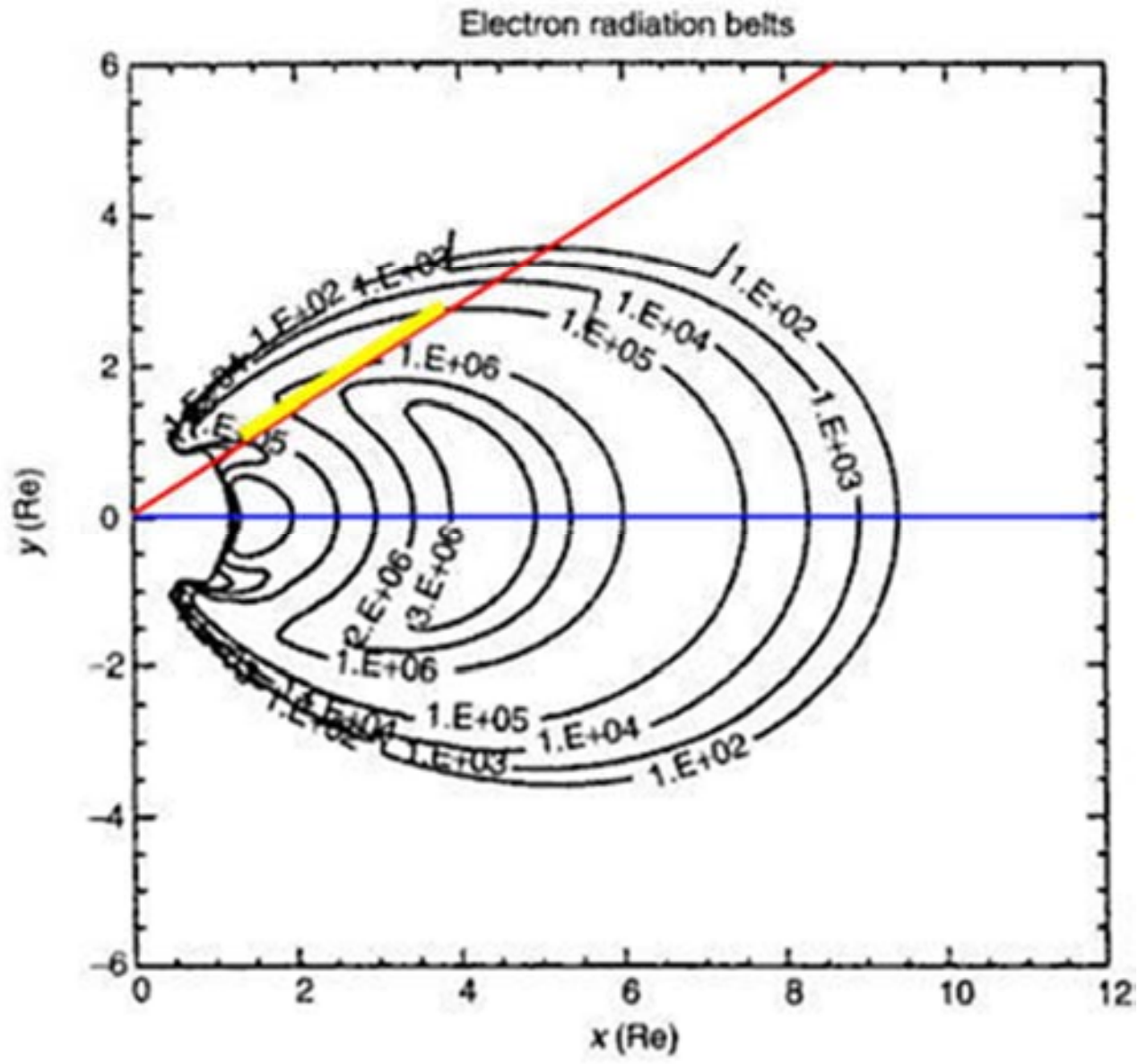

***Figure 8.*** *The Outer Part of the VAB: The Electron Radiation Belt.*

Figure 8 shows that the exit path in the electron belt is in an area where already small changes (upwards or downwards) have a large effect.

If therefore the Apollo 11 flight path were located slightly above the shown line then the dose calculation along the red path would yield a too high dose. For this reason I present in the next chapter a total dose calculation with the flight path as it is described in the Mission Report [10]. So I make here no estimation of the total dose along such a straight flight path.

## 4. Estimation of the Radiation Dose According to the Apollo 11 Flight Path

The basic assumptions for the total dose calculation are unchanged with respect to the previous chapter. Also here a value is calculated which could have been achieved under the most favourable radiation conditions.

Here I use the exact flight path. It is described in chapter 7. The flight to the Moon and the flight back to the Earth are calculated separately.

Figure 9 and Figure 10 show the path through the radiation belt, the inner part with mainly protons and the outer part with mainly electrons. On the top there is the path to the Moon, on the bottom the return to the Earth. The small red circles are points of the trajectory as they are given in the "Apollo 11 Mission Report" [10]. Additionally in Figure 10 the associated manoeuvres are indicated. Surprisingly the flipping manoeuvre of the Command and Service Module (CSM) (between CM/S-IVB Separation and Docking) is in the area of the maximum radiation.

The small blue circles are points which have been used to draw the trajectory in Figure 9 and Figure 10. These circles are often entry or exit points of radiation zones (see Table 2 below).

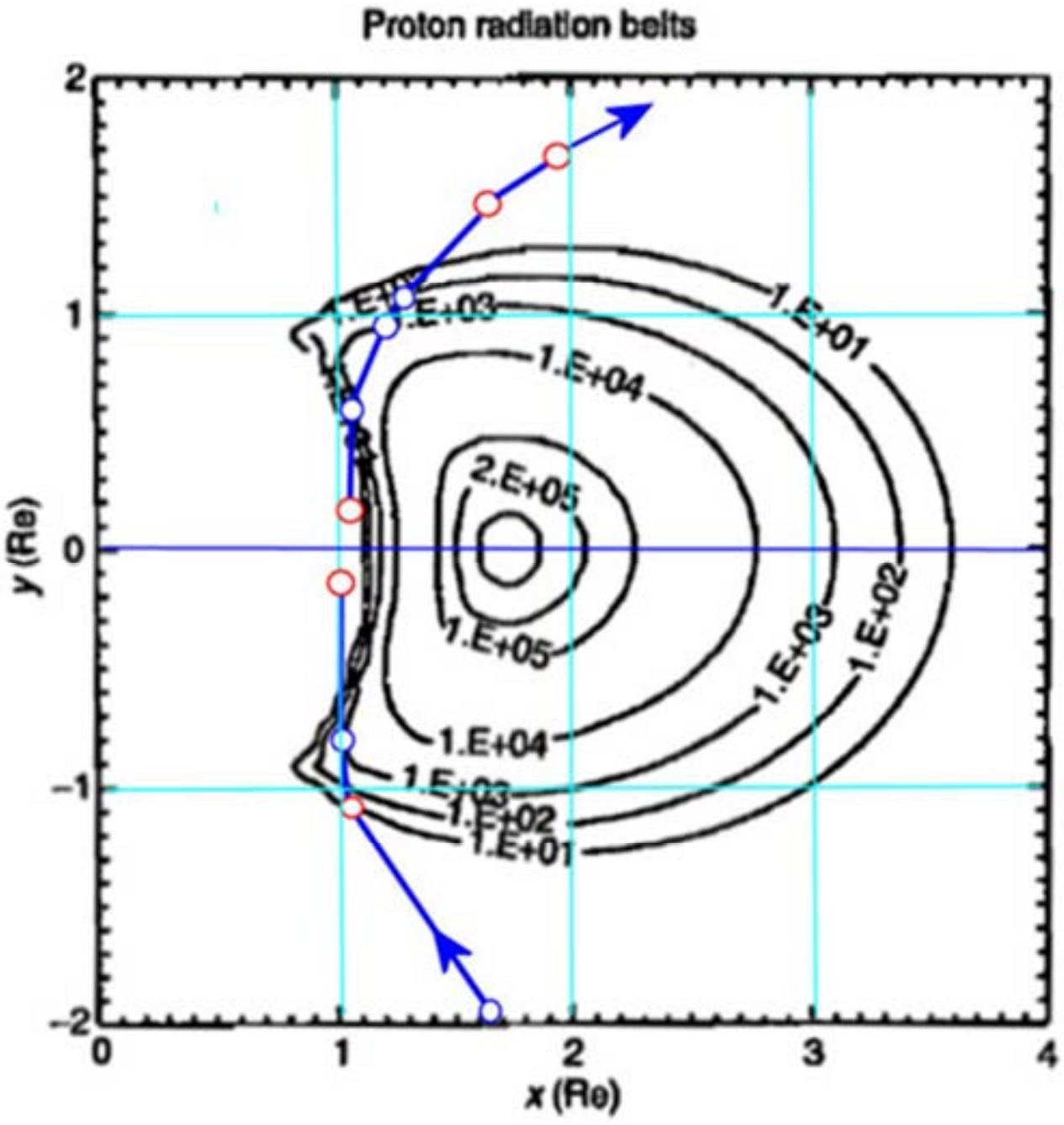

***Figure 9.*** *Flight Path through the Proton Van Allen Radiation Belt.*

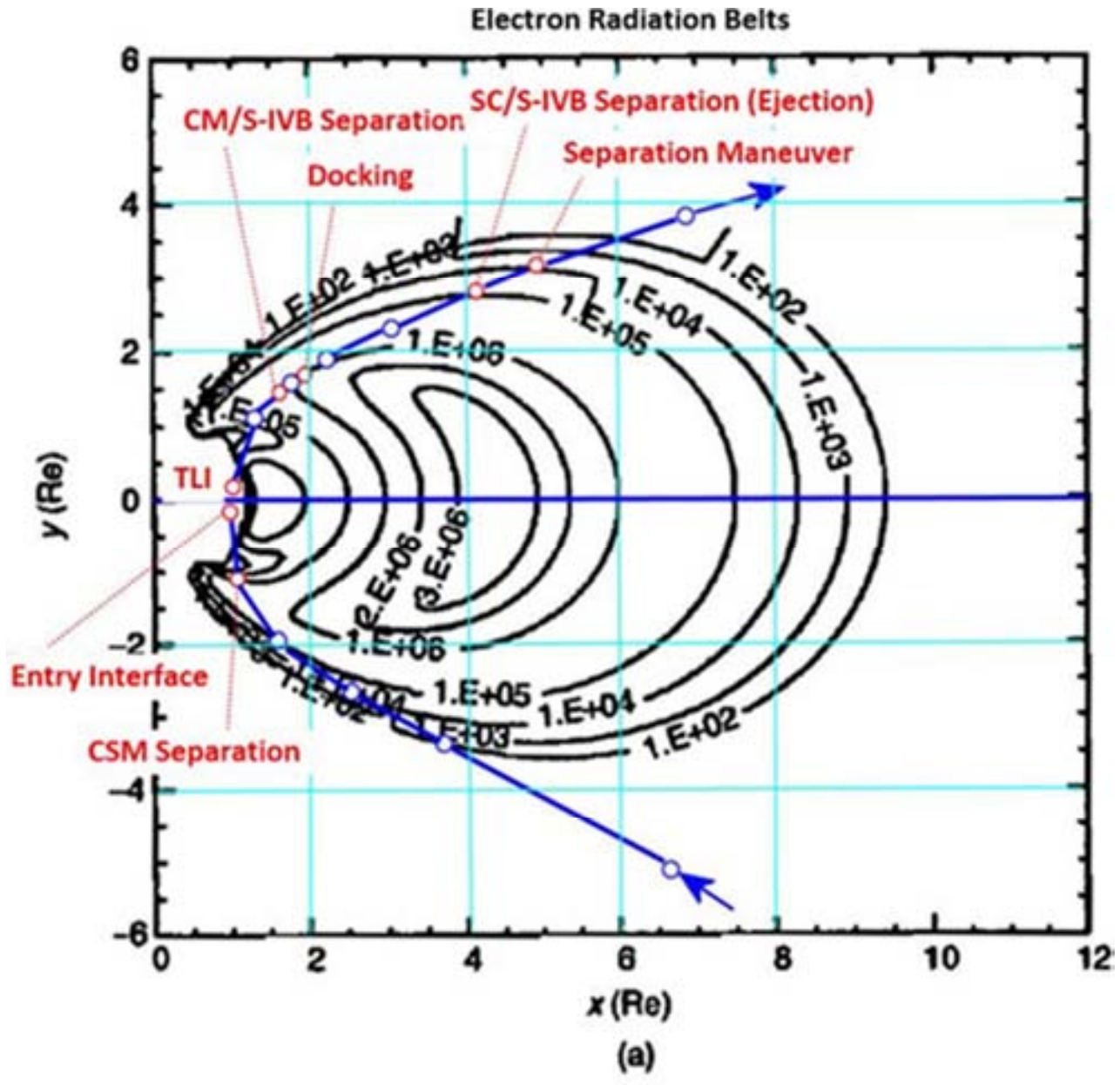

***Figure 10.*** *Flight Path through the Electron Van Allen Radiation Belt.*

Comparing these two figures with the ones in the previous chapter one recognises that the flight path obviously crosses the Van Allen radiation belt quite exactly during its maximum inclination. The trajectory is slightly above the ecliptic and circumvents the central region even better.

Also the return path is more favourable. Here the fact helps that the Moon was at the time of the departure from the Moon already 2° below the ecliptic.

With this data and the knowledge of the exact time between the different points of the trajectory the total radiation dose can be determined.

The total dose is calculated analogously to the level curves of the high energy protons and electrons, i.e. the maximum dose in the centre of the two belts is reduced according to the level curves. This procedure is justified on the one hand that the total dose is proportional to the number of the (high energy) particles, on the other hand in Figures 4 and 5 (annual dose) the total dose decreases along the x-axis in the same extent as the level curves in Figures 2 and 3, e.g. about a factor of 3 from 5·Re to 6·Re.

The calculation of the total dose is demonstrated for a shielding of 4 mm aluminium. The maximum intensities are according to Figures 4 and 5: for the proton belt this is $4.1\cdot10^5$ rad/year or 465 mSv/h and for the electron belt $3.1\cdot10^5$ rad/year or 355 mSv/h.

Table 2 shows the determination of the total radiation dose. I counted only the permanent available radiation in the Van Allen radiation belt, i.e. no solar radiation.

For the intensity the lowest value of the respective zone is used. From the maximum of all zones ($3\cdot10^5$ for protons and $3\cdot10^6$ for electrons) to the next lower zone (the lower end is a power of 10) the maximum value is reduced by a factor of 3, then for every lower zone by another factor of 10.

***Table 2.*** *Determination of the Radiation Dose for 4mm Al Shielding through the VAB.*

| | Zone | Time in Zone | Dose Calculation | Dose |
|---|---|---|---|---|
| Path to the Moon | 1E3 … 1E4 $p^+$ | 400s ≈ 7min | 400s * (1/300)*465mSv/h | >0.2 mSv |
| | 1E4 … 1E5 $e^-$ | ≈0s | ≈0 mSv | ≈0 mSv |
| | 1E5 … 1E6 $e^-$ | 800s ≈ 13min | 800s * (1/30)*355mSv/h | >2.6 mSv |
| | >(≈) 1E6 $e^-$ | 700s ≈ 12min | 700s * (1/3)*355mSv/h | >(≈) 23.0 mSv |
| | 1E6 … 1E5 $e^-$ | 2700s=45min | 2700s * (1/30)*355mSv/h | >8.9 mSv |
| | 1E5 … 1E4 $e^-$ | 1383s ≈ 23min | 1383s * (1/300)*355mSv/h | >0.5 mSv |
| Total Outward | | | | >35.2 mSv |
| Return Path | 1E4 … 1E5 $e^-$ | 1500s=25min | 1500s * (1/300)*355mSv/h | >0.5 mSv |
| | 1E5 … 1E6 $e^-$ | 1200s=20min | 1200s * (1/30)*355mSv/h | >3.9 mSv |
| | 1E5 … 1E4 $e^-$ | ≈0s | ≈0 mSv | ≈0 mSv |
| Total Return | | | | >4.4 mSv |
| Total Resulting Dose | | | | >39.6 mSv<br>>39.6 mGy |
| Apollo 11 Mission Dose | | | | 1.8 mGy |

The total dose depends on the shielding thickness (and material). From the total dose calculation [3] there result also dose values for further shield thickness'.

Table 3 shows the total dose in dependency of the shielding. The dose @ 4 mm Alu shielding is 39.6 mSv as in Table 2.

The mission dose value of 1.8 mGy corresponds in Table 3 to a shielding of slightly more than 7 mm. For comparison: according to Table 1 which is calculated for 1969 a shielding of greater 20 mm would have been necessary – even for low particle fluences from the Sun.

Such heavy radiation shielding is not necessary for flights in equatorial Low Earth Orbits up to 500 km, but for flights higher than 1'000 km altitude it is crucial.

Assessment of the radiation shielding of the CM:

The inner structure of the Command Module consists of an aluminium honeycomb sandwich bonded between sheet aluminium alloy. The outer structure, the heat shield, is made of steel honeycombs. [15, 16] With this construction technique one can get a high stability with little material.

Areas with low radiation shielding excessively reduce the total shielding effect. In summary the above required 7 mm shielding may look about realistic.

After my analyses I found a NASA web site [18] which is quite well in line with my findings: NASA assumed for the CM 2 g/cm$^2$ (7.4 mm aluminium) and for the space suit 0.17 g/cm$^2$ (0.6 mm aluminium).

***Table 3.*** *Total Lunar Mission Dose as a Function of the Shielding Thickness (with a quiet Sun, i.e. Radiation within the VAB only).*

| Shielding [mm Aluminium] | Total Mission Dose [mSv] |
|---|---|
| 0.05 | 9'175.1 |
| 0.1 | 6'297.3 |
| 0.2 | 4'105.4 |
| 0.3 | 2'999.9 |
| 0.4 | 2'308.4 |
| 0.5 | 1'835.1 |
| 0.6 | 1'503.3 |
| 0.8 | 1'087.3 (109 rad) |
| 1 | 835.5 |
| 1.5 | 485.4 |
| 2 | 290.1 |
| 2.5 | 173.1 |
| 3 | 104.0 (10 rad) |
| 4 | 39.6 |
| 5 | 15.5 |
| 6 | 5.9 |
| 7 | 2.1 |
| 8 | 0.8 |
| 9 | 0.4 |
| 10 | 0.2 |
| 12 | 0.2 |
| 14 | 0.1 |
| 16 | 0.1 |
| 18 | 0.1 |
| 20 | 0.1 |

## 5. Prediction of the Radiation

Here I present predictions for radiation doses for Moon and Mars flights. For a good overview I assume only a shielding of 7 mm for the space craft and of 1 mm for the space suit.

The flight from Earth to Mars lasts with the current technology about 8 months [17]. The dose rate between Earth and Mars is assumed to be constant, so the travel dose is proportional to the flight time: 8 months (one way) versus 7.5d (as used in Table 1): the travel dose in Table 1 has to be multiplied with 8·30/7.5 which is 32.

Flight to the Moon, in a year of a solar maximum (e.g. 1969):
Van Allen belt passage (2x): 2.1 mSv [Table 3]
Journey to the Moon and back: 361 mSv [Table 1]
2 h on the lunar surface: 43 mSv [Table 1]
Total dose: 406.1 mSv (41 rad)
Flight to the Moon, in a year of a solar minimum (e.g. 1996):
Van Allen belt passage (2x): 2.1 mSv [Table 3]
Journey to the Moon and back: 0 mSv
2 h on the lunar surface: 0 mSv
Total dose: 2.1 mSv
Flight to Mars, in a year of a solar maximum (e.g. 1969):
Van Allen belt passage (1x): 1 mSv [Table 3]
Journey to Mars (one way): 11’552 mSv
Total dose: 11’553 mSv (1.2 krad)
Flight to Mars, in a year of a solar minimum (e.g. 1996):
Van Allen belt passage (1x): 1 mSv [Table 3]
Journey to Mars (one way): 0 mSv[Note 1]
Total dose: 1 mSv

Note 1: This value is based on the default model for solar protons (ESP). Another model, the King model, predicts 95 mSv.

The above data show the span of the possible radiation. Solar flares can produce even stronger radiation rates than these values which are based on a one year average. Further radiation sources as galactic or cosmic heavy ions have not been considered. All this makes a manned space flight outside of the Van Allen radiation belt to a not calculable risk.

Electronics has generally a design margin of 2, i.e. it is tested to twice the expected radiation dose. I would expect a margin greater than 2 for manned space flight.

The above radiation levels are small for electronics. Even commercial electronics starts to degrade only after 1...10 krad. Humans are more susceptible, as shown in the next chapter.

The risk for electronics components is no principal problem. A 90% probability for a success may be OK for a robotic mission with a pioneer character. But I doubt whether this would be OK for a manned mission as well.

As soon as a space craft is on the way to Moon or Mars there is no possibility to return. On the surface of the Moon the astronauts can plan extravehicular activities depending on the space weather. But if a strong solar flare event occurs in the direction of the space craft then the crew is lost.

# 6. Effects of Radiation

The impact of radiation is shown in the following table [5]. Similar data can be found in [22].

***Table 4.*** *Effects of Radiation on Humans.*

| Dose | Radiation effect |
|---|---|
| 0 to 0.5 Sv (0 to 50 rad) | Without greater diagnostic effort no immediate disadvantageous effects noticeable, but degradation of the immune system |
| 0.5 to 1 Sv (50 to 100 rad) | Changes in the blood picture, erythema, sporadic nausea, vomiting, very rare events of death |
| 1 to 2 Sv (100 to 200 rad) | Disadvantageous effects on the bone marrow, vomiting, nausea, bad general condition, about 20% mortality |
| greater 4 Sv (greater 400 rad) | Severe constraints of the general condition and heavy disturbances on the sanguification. The disposition to infections is strongly increased, 50% mortality |
| greater 6 Sv (greater 600 rad) | Besides the named heavy disturbances there appear gastrointestinal symptoms. The survival rate is very low |
| over 7 Sv (over 700 rad) | Almost 100% mortality |
| over 10 Sv (over 1 krad) | Additional damage of the central nervous system, up to paralyses |
| over 100 Sv (over 10 krad) | Fast death caused by a malfunction of the central nervous system (sudden death) |

The natural annual dose is around 2.5 - 4.5 mSv.

The optimistically determined mission dose of the previous chapter is according to the above table well in the save area.

The maximum operational dose limit for each of the Apollo missions was according to [13] *“set at 400 rad to skin and 50 rad to the blood-forming organs… In the heavy, well shielded Command Module, even* during *one of the largest solar-particle event series… the crewmen would have received a dose of 360 rad to their skin and 35 rad to their blood-forming organs (bones and spleen).”*

This estimation of the received radiation dose (received during a large solar particle event) fits perfectly to the maximum dose limits, but it is in contradiction to the physical effect of radiation and shielding: the radiation which passes the cover of the CM is a penetrating radiation[Note 2]. The additional shielding effect of the skin is then negligible, so that all organs receive the same dose.

Note 2: The penetrating radiation consists in this context of proton radiation and of Bremsstrahlung, which becomes manifest in gamma radiation.

# 7. Determination of the Flight Path

Several points of the flight path of the Apollo 11 mission are exactly indicated in the Mission Report [10] including the velocity vector. With this the flight path can be calculated with help of numerical simulation. So in the end all data including time stamps are available.

All reported points almost exactly correspond with the simulation. The used curves can therefore be regarded as perfect for this consideration.

For the numerical simulation the following differential equation is integrated.

$$\ddot{\underline{r}} = -\frac{\Gamma\cdot M}{|\underline{r}|^3}\cdot \underline{r} \qquad (1)$$

$r$ is the vector from the centre of the Earth to the space craft; $\Gamma$ is the gravitational constant ($6.674\cdot10^{-11}$ $m^3/(kg\cdot s^2)$); M is the mass of the Earth ($5.976\cdot10^{24}$ kg) and $\ddot{\underline{r}}$ is the 2nd time derivative of $\underline{r}$, i.e. the acceleration vector.

The reference system is Earth fixed (not rotating with the Earth, i.e. inertial): the origin is the centre of the Earth, the x-axis in the direction of the vernal equinox, the z-axis=Earth axis in the direction of the North Pole and the y-axis results from the right-handed system.

The position of the Earth axis relative to the Moon and the position of the Earth on its orbit are shown in Figure 12.

Figure 12 shows the constellation at TLI (Translunar Injection), i.e. after the acceleration phase behind the Earth: from this point in time the flight goes “in free fall” in direction Moon. The direction of the Moon corresponds already to the one of the arrival in the lunar orbit of Apollo 11, at 174° [9, 10] (ecliptic) longitude (from vernal equinox ν).

The Sun is shown in yellow in the middle of Figure 12. On the left there is once again the Earth on March 21, i.e. at equinox. The inclination of the equator is indicated as well.

The geomagnetic pole was 1969 at (78.5°N, 70°W) [8]. Its direction from the North Pole is indicated.

TLI was behind the Earth, seen from the Moon: (10°N, 165°W) [10]. The direction is also indicated.

For a better imagination of the trajectory one can take the logo of the Apollo Flight Journal [12]. I have complemented it with directional arrows in Figure 11.

The main results of the trajectory calculation have already been worked into chapter 4, specifically in Figures 9 and 10.

For a better understanding of the flight path I present several additional curves to you: first in Figures 13 and 14 the flight path, as it already has been shown in Figure 9 and Figure 10. The points which have been used for the dose calculation in the electron belt and which are marked with blue circles in Figure 10 are also here indicated with blue circles.

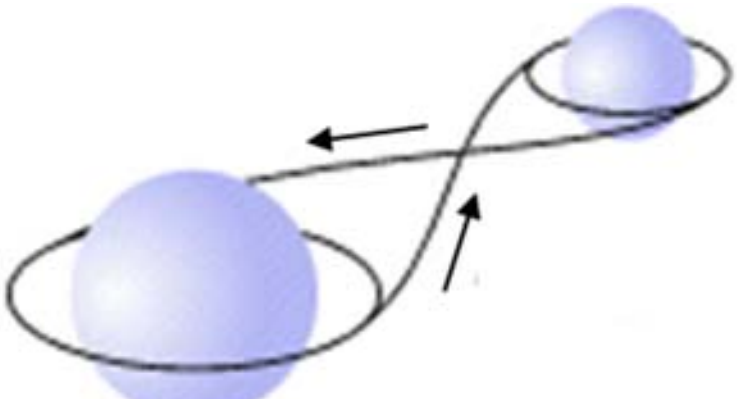

***Figure 11.*** *Logo of the Apollo 11 Flight Journal with the Flight Path.*

The Figures 15-20 show the flight path in the equatorial and in the ecliptic reference system. The latter corresponds probably the best with our imagination: the Moon was at the approach about in the ecliptic, at the departure it was 2° below the ecliptic.

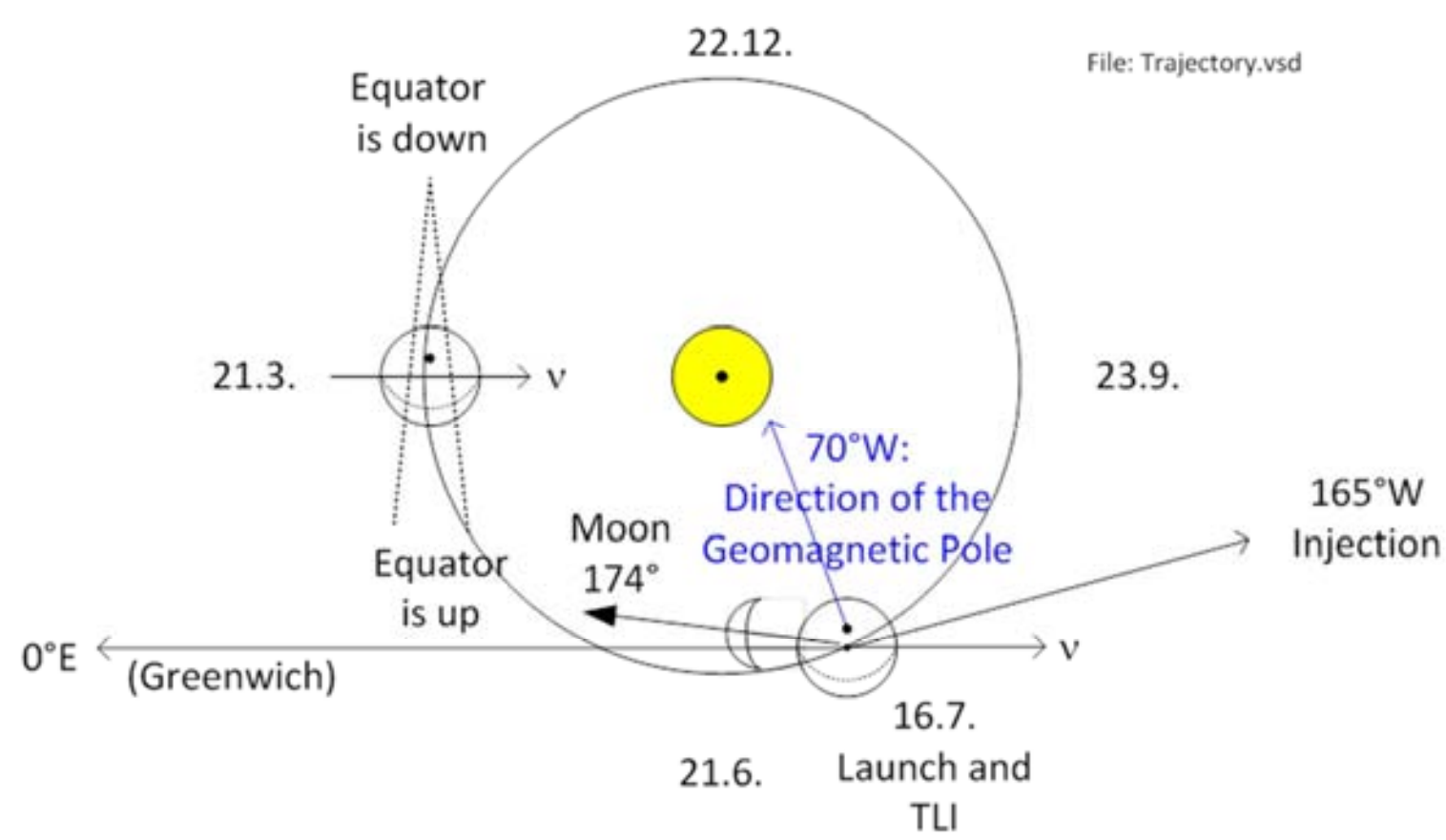

***Figure 12.*** *Constellation of Sun, Earth and Moon at TLI of Apollo 11.*

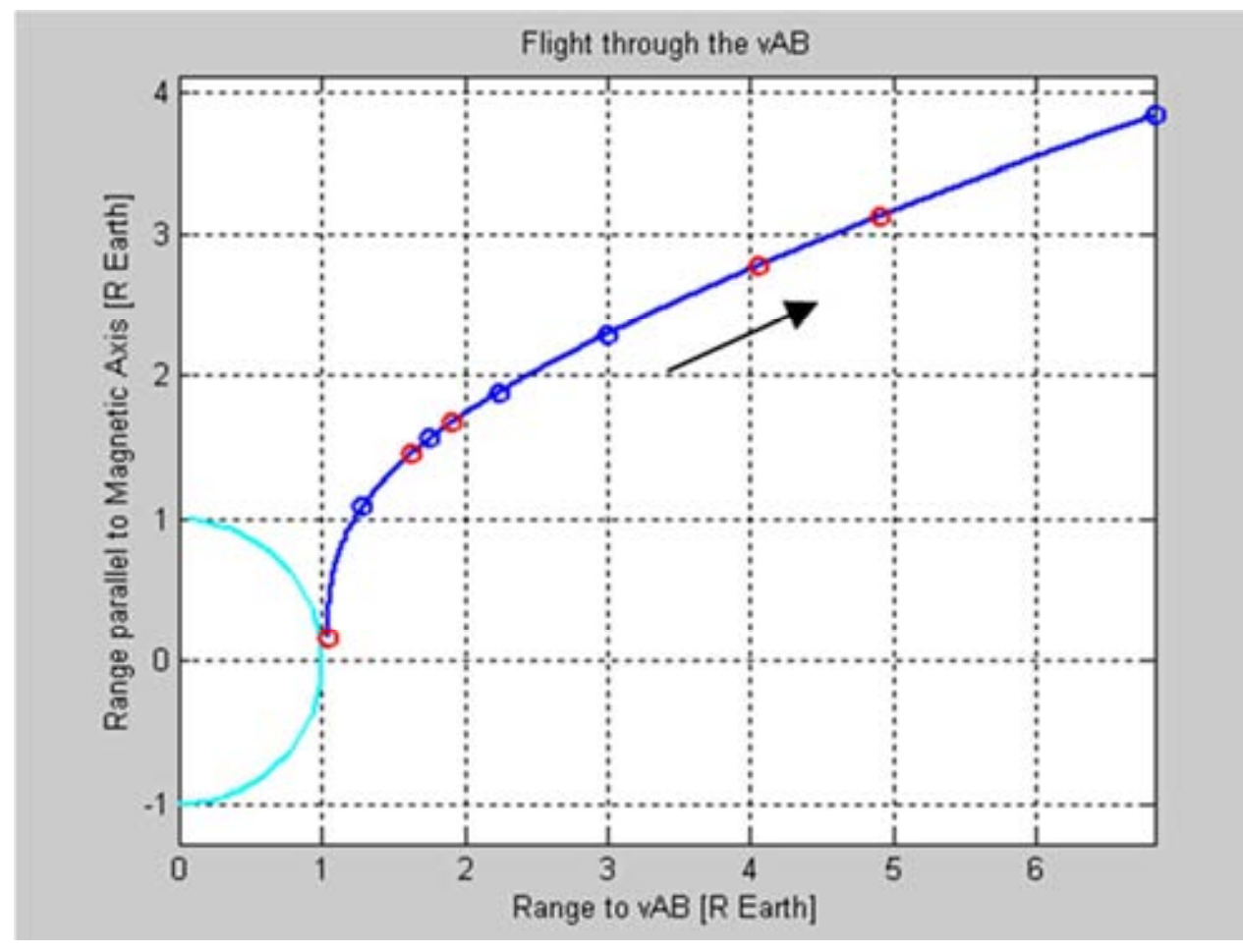

***Figure 13.*** *Flight Path in the Geomagnetic Reference System.*

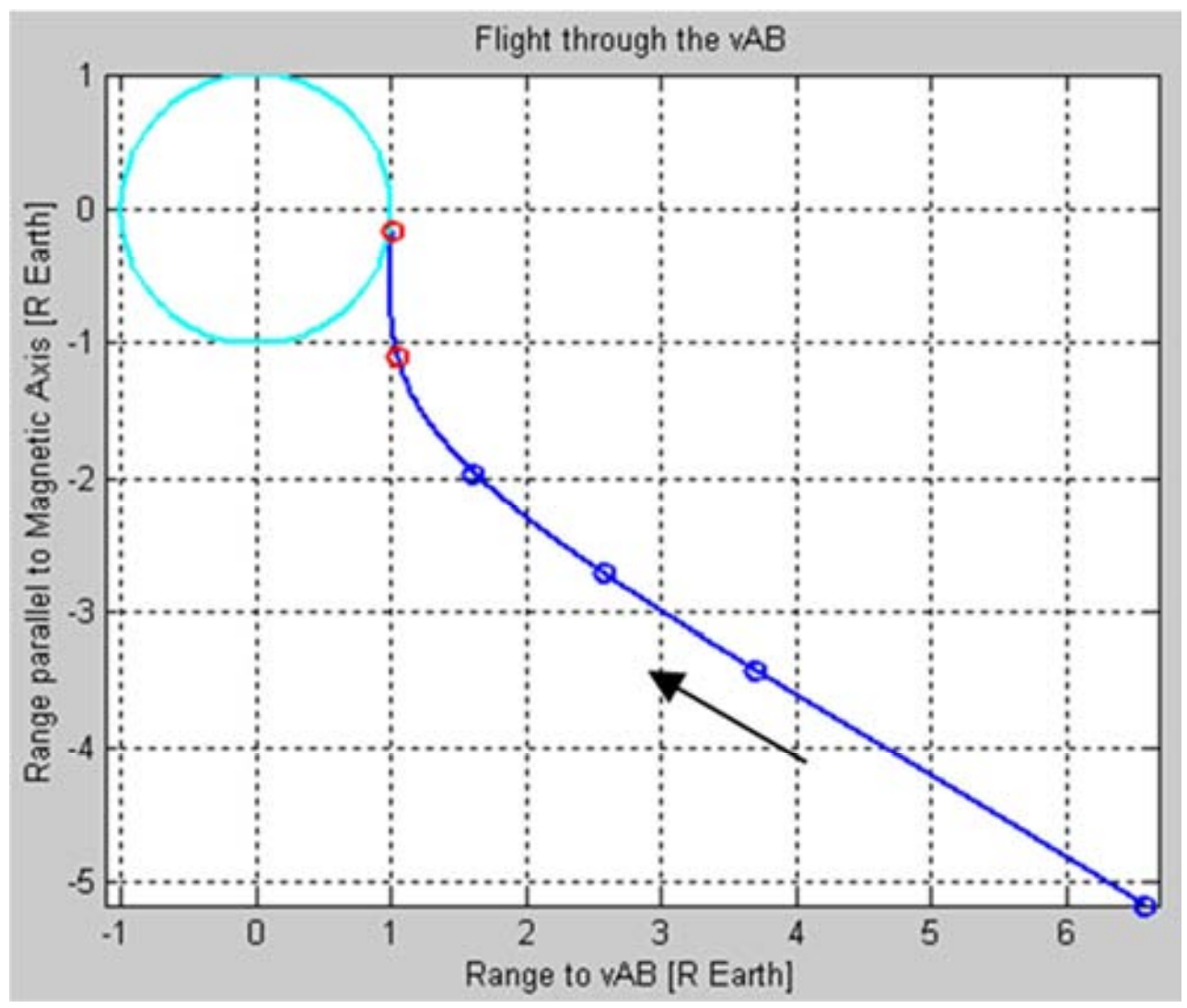

***Figure 14.*** *Flight Path in the Geomagnetic Reference System.*

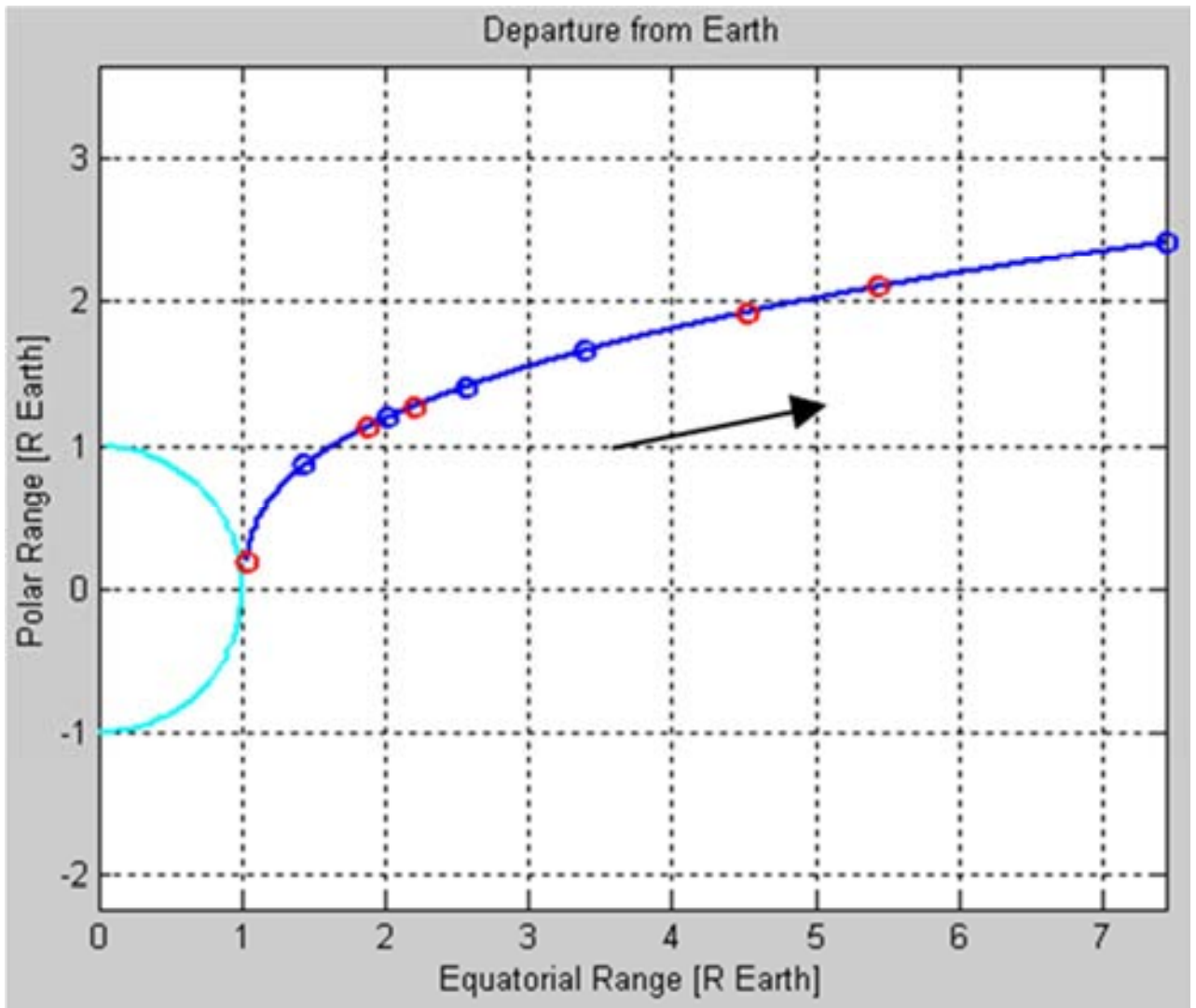

***Figure 15.*** *Flight Path in the Equatorial Reference System.*

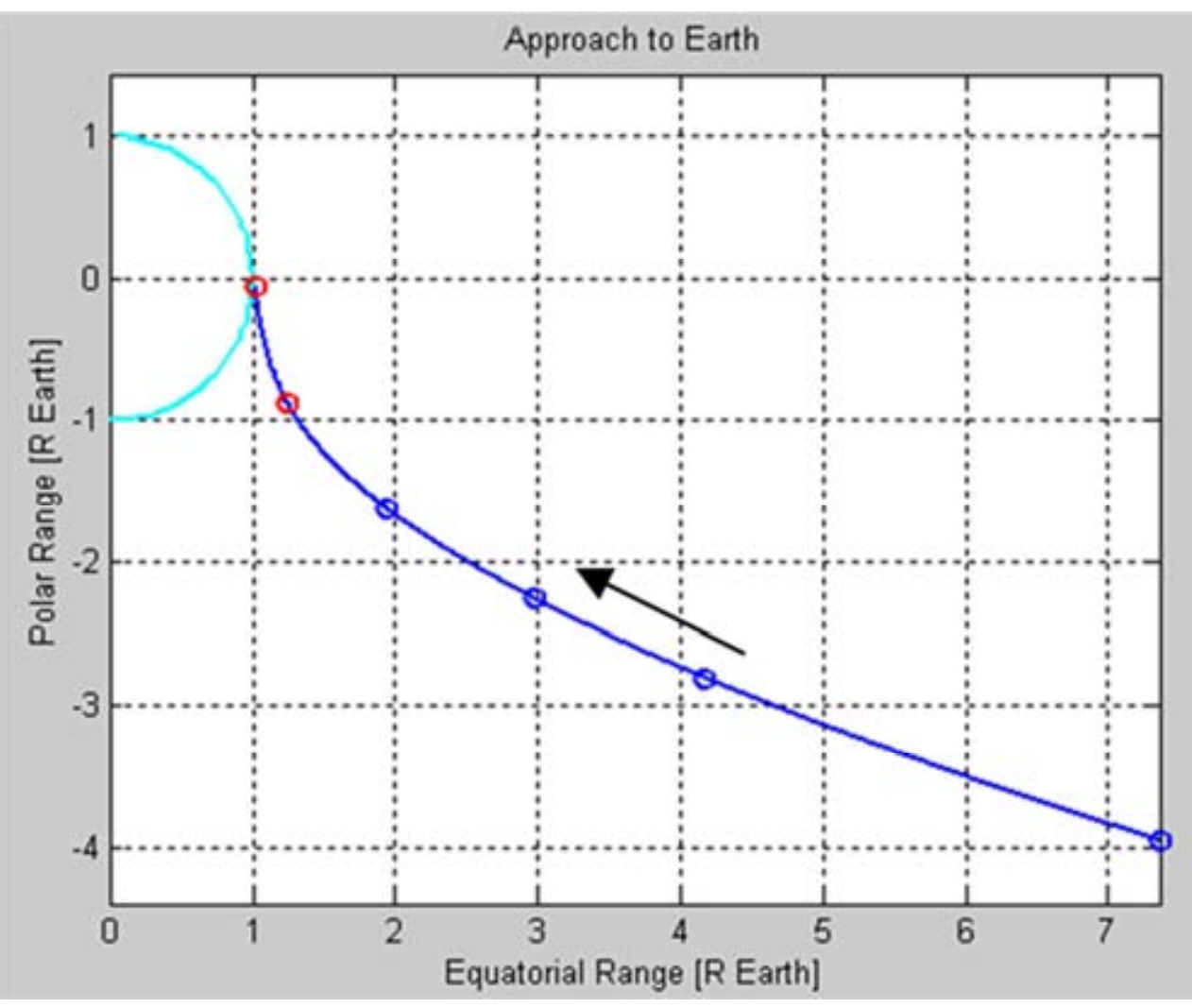

***Figure 16.*** *Flight Path in the Equatorial Reference System.*

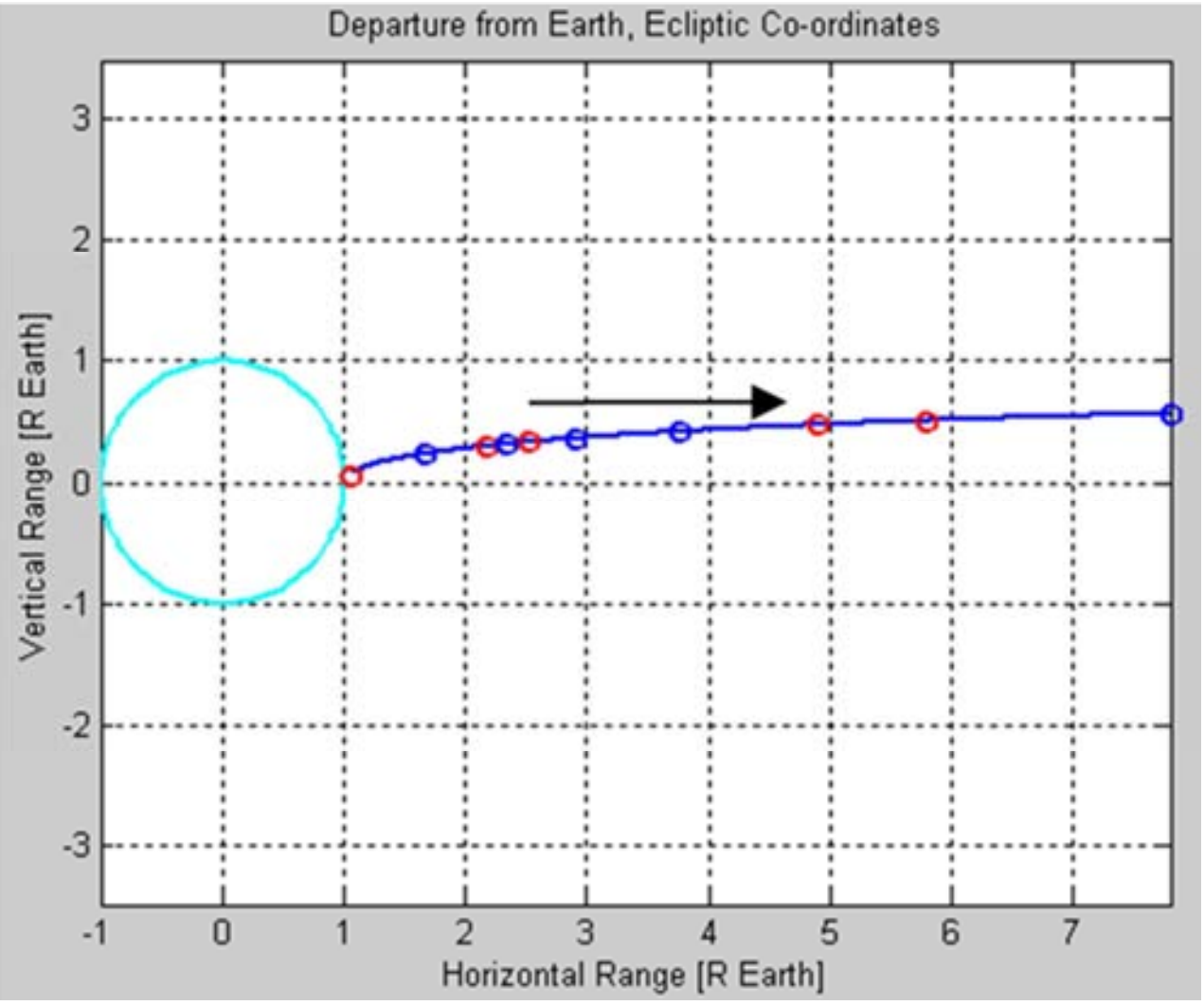

***Figure 17.*** *Flight Path in the Ecliptic Reference System.*

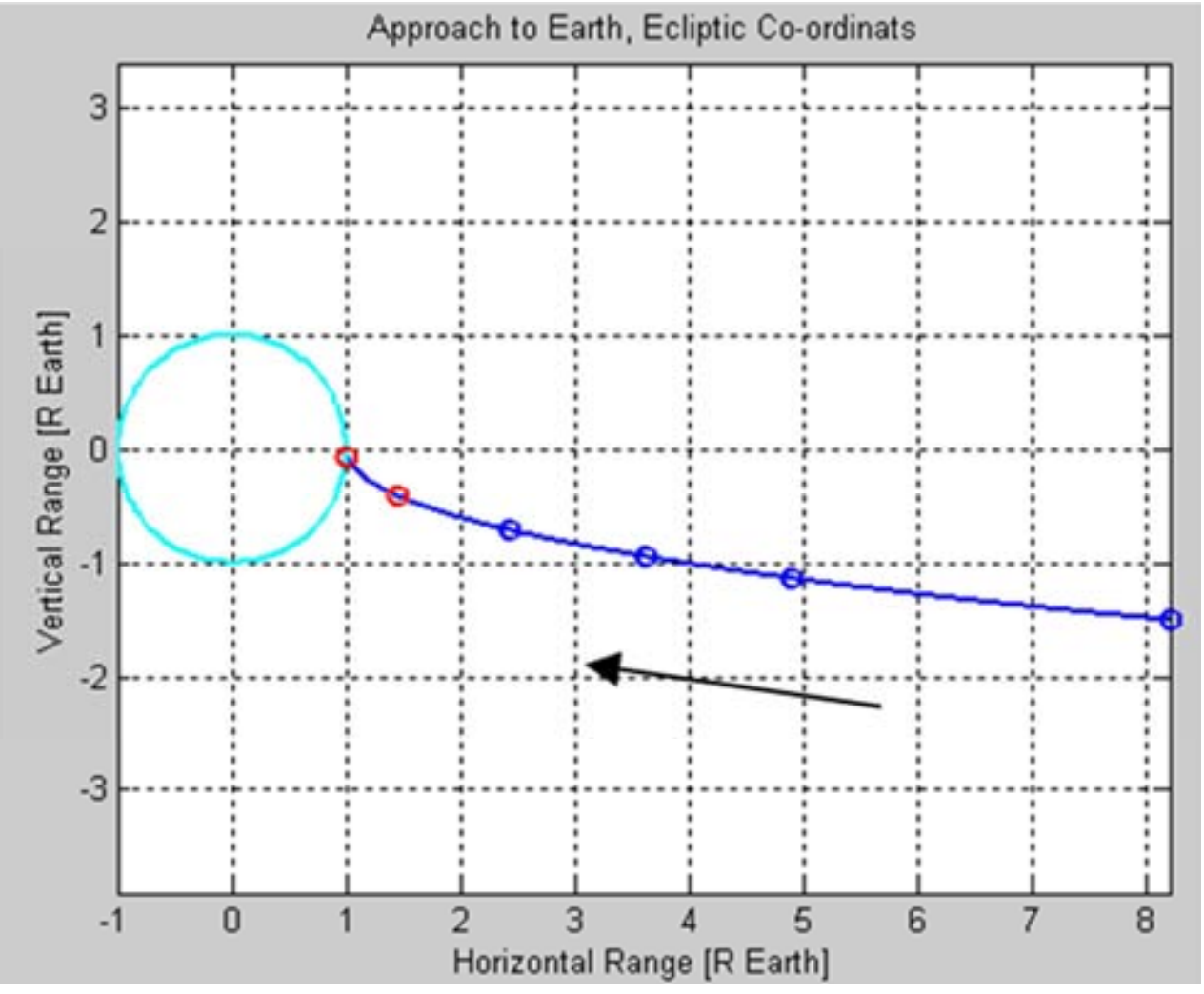

***Figure 18.*** *Flight Path in the Ecliptic Reference System.*

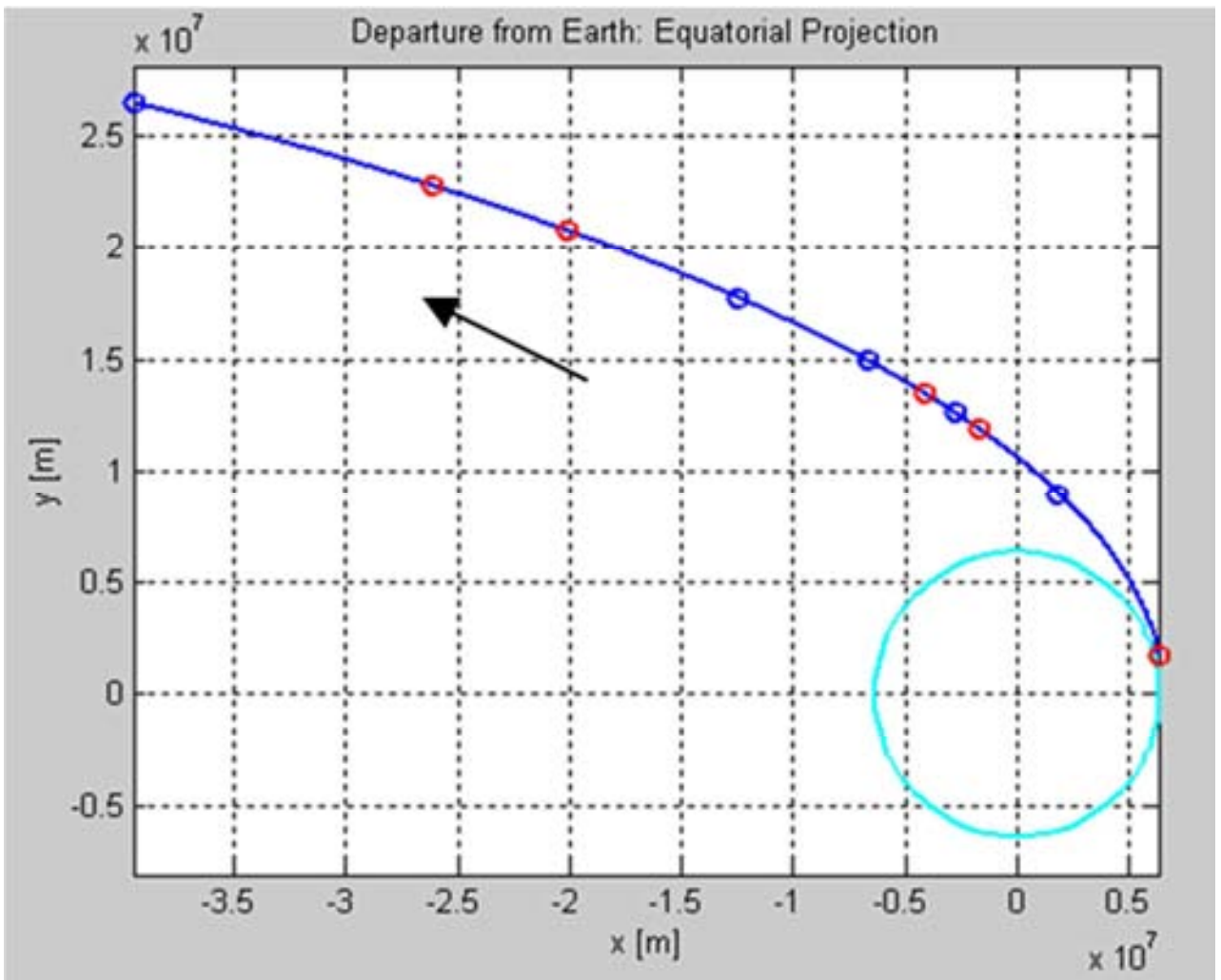

***Figure 19.*** *Projection of the Flight Path in the Equatorial Plane.*

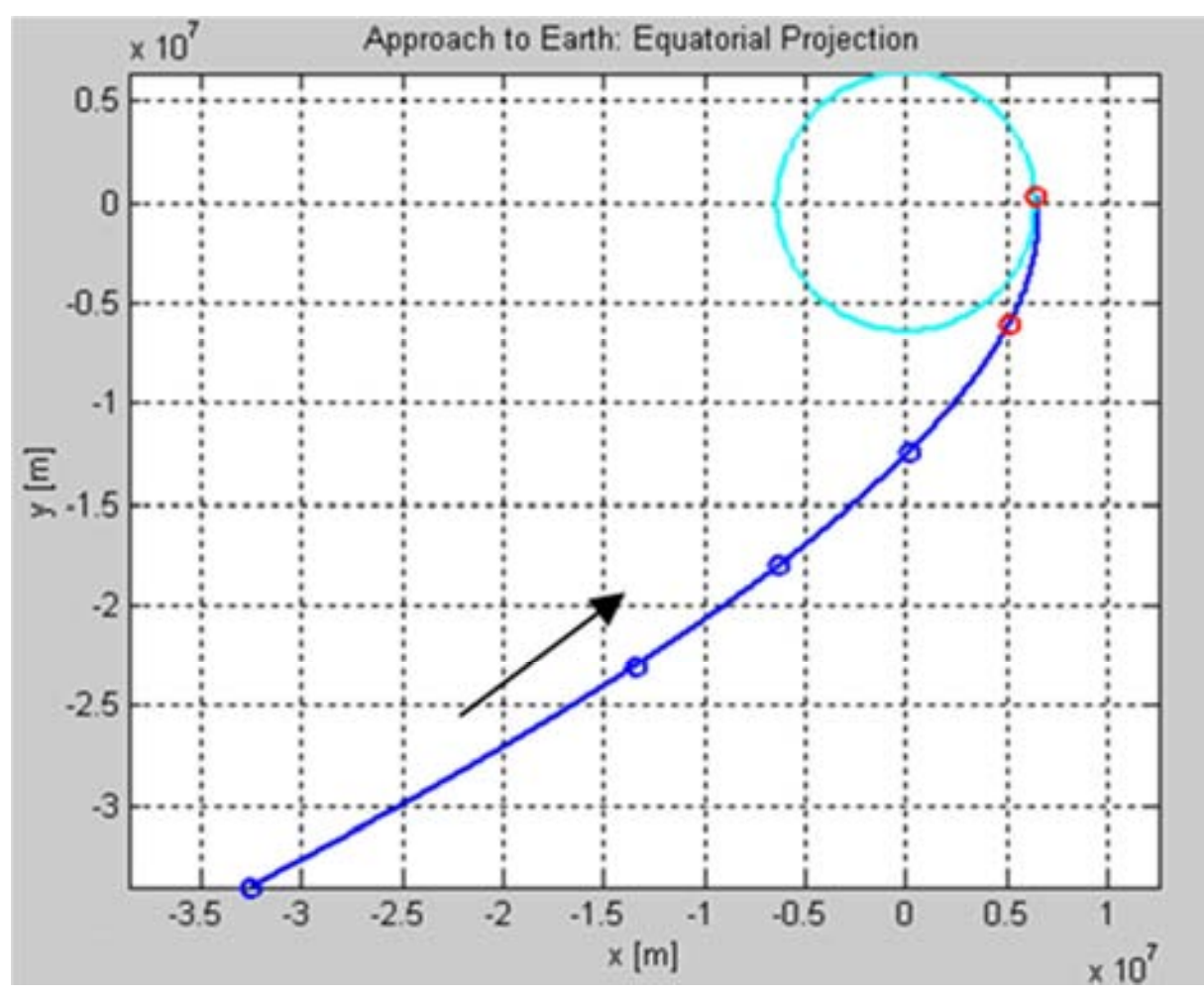

***Figure 20.*** *Projection of the Flight Path in the Equatorial Plane.*

According to Figures 21-24 it looks as if the space craft during the departure (Figures 21 and 23) flew like a high jumper over the Van Allen radiation belt; during the return flight on Figures 22 and 24 we see the same behaviour but down under. The trajectory to the waxing Moon was very well tuned to avoid the central part of the Van Allen radiation belt.

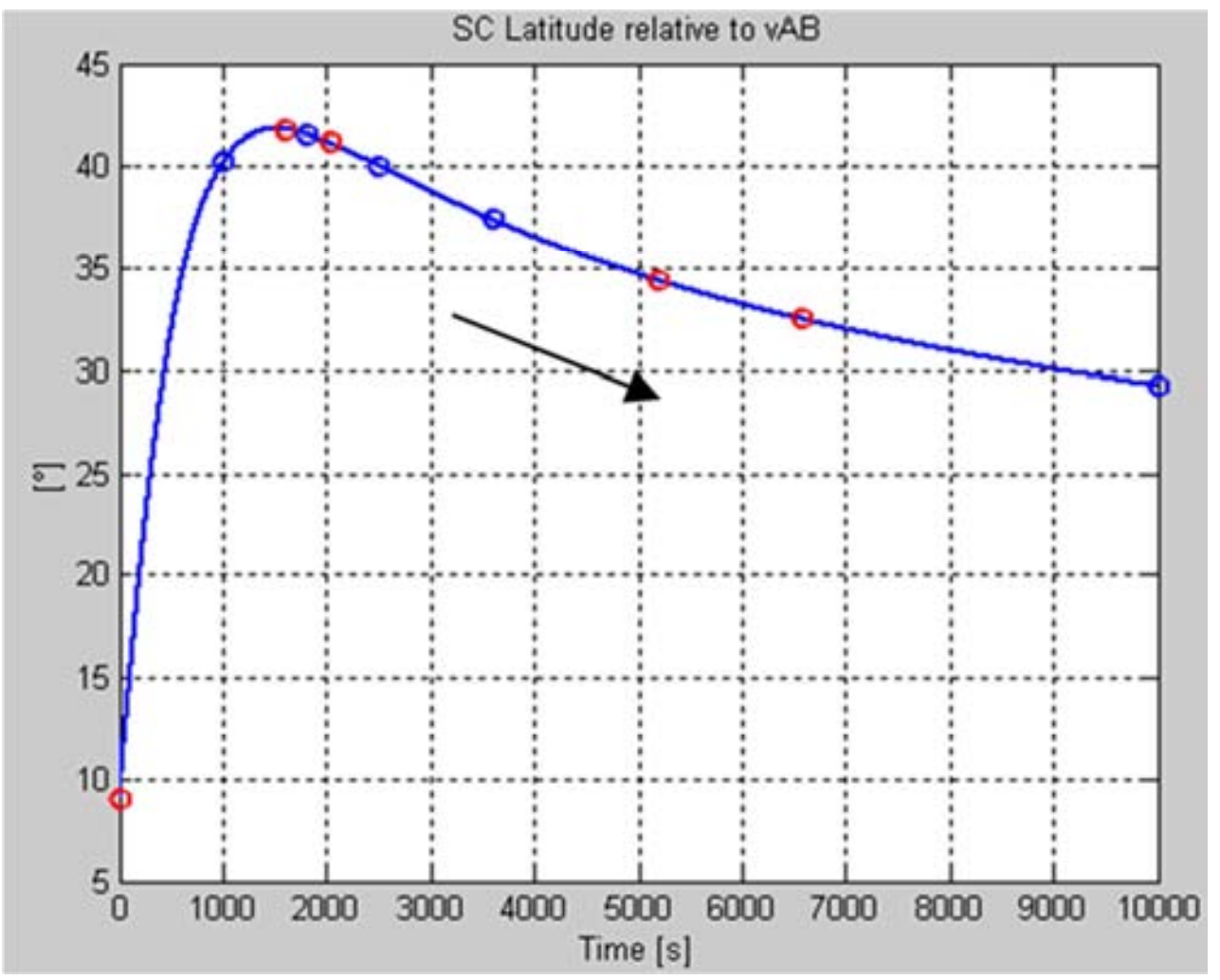

***Figure 21.*** *Latitude in the Geomagnetic Reference System.*

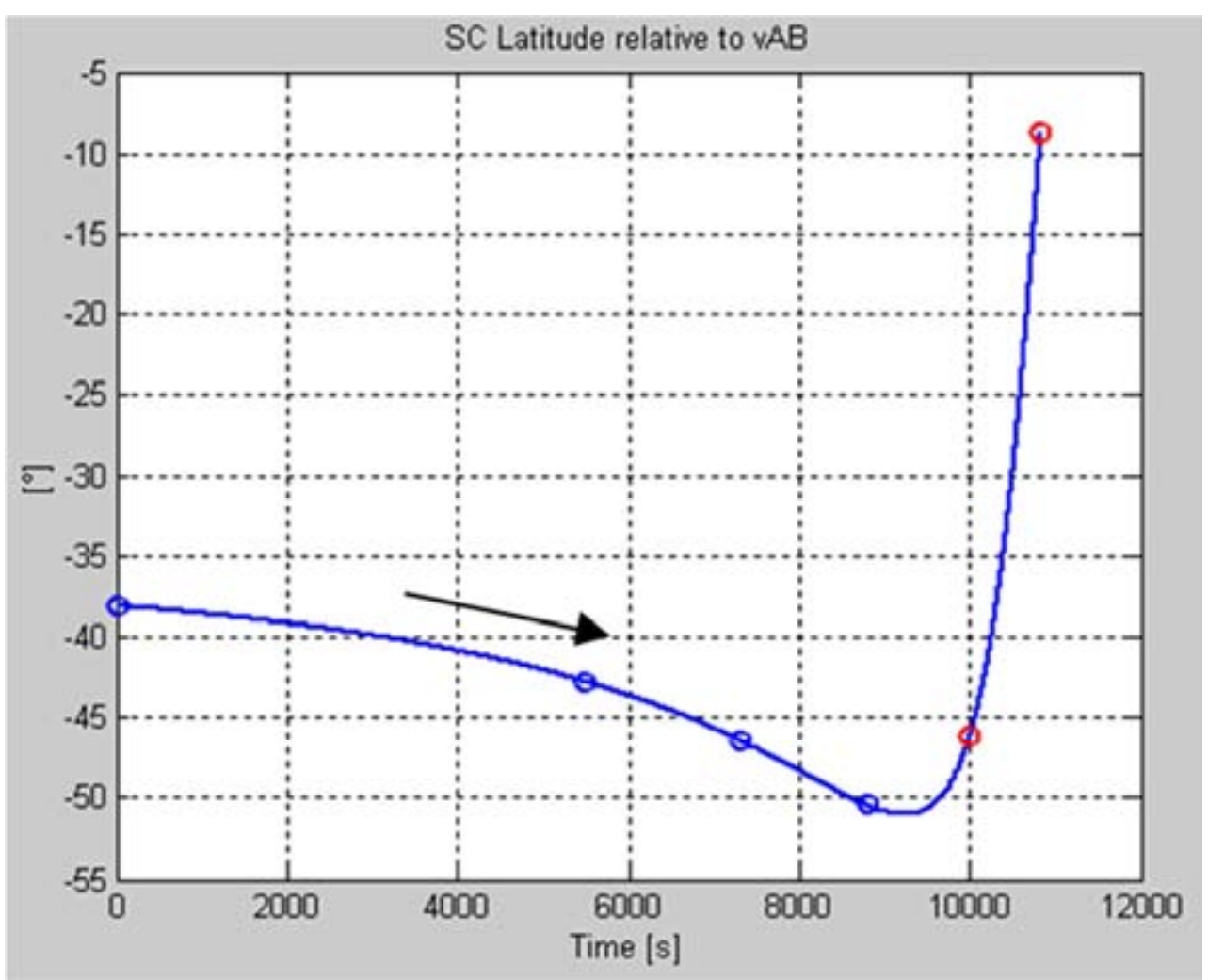

***Figure 22.*** *Latitude in the Geomagnetic Reference System.*

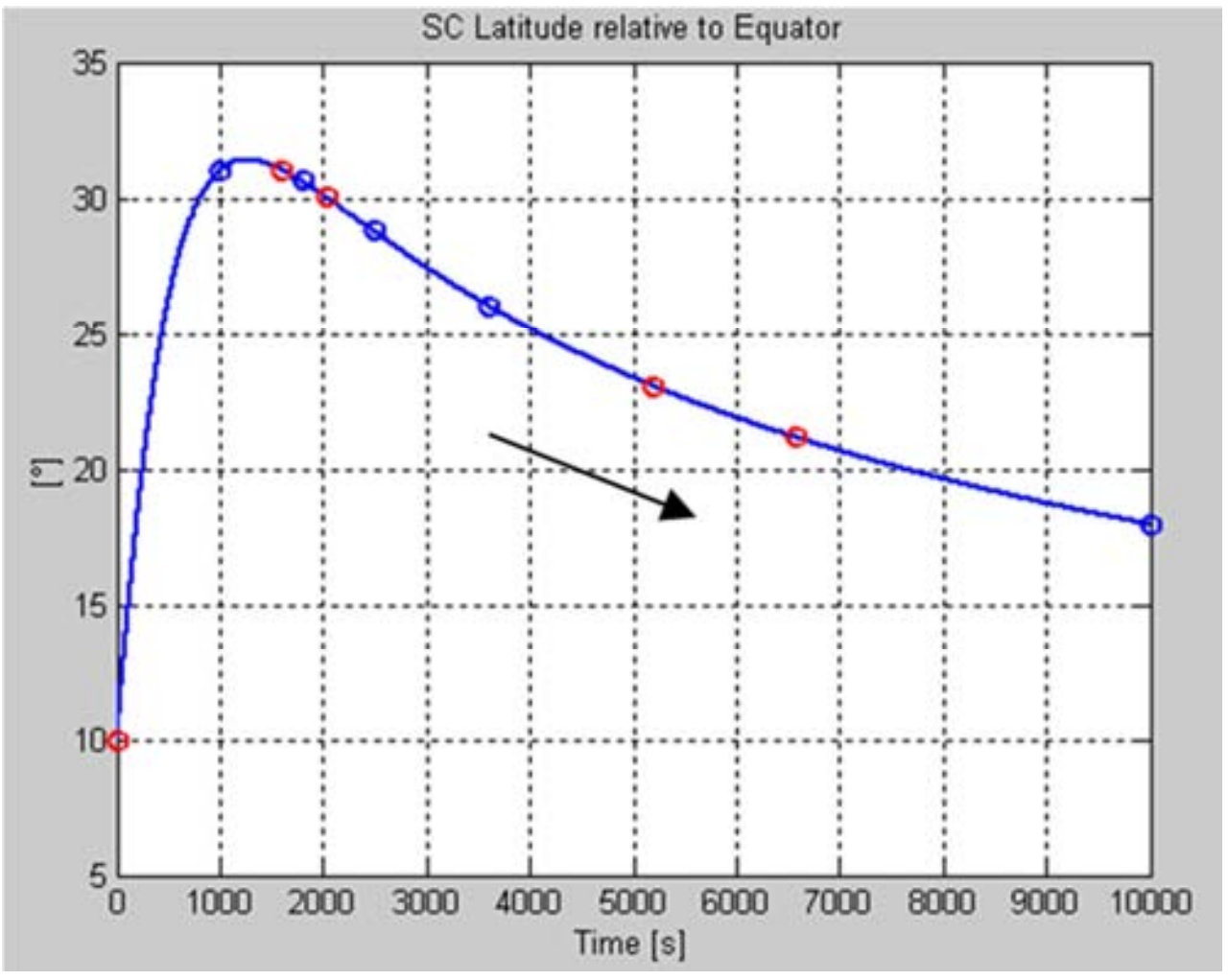

***Figure 23.*** *Latitude (in the Equatorial Reference System).*

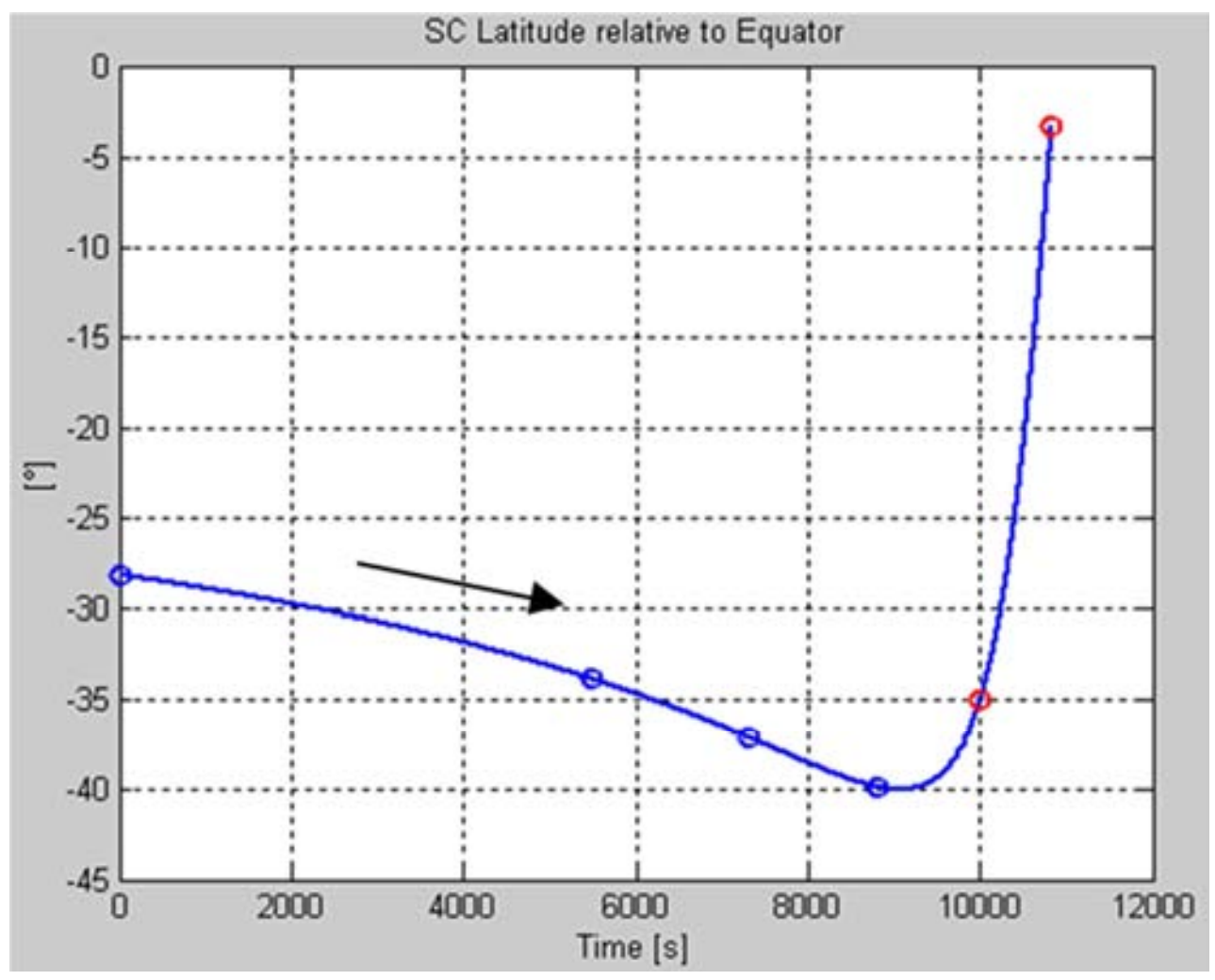

***Figure 24.*** *Latitude (in the Equatorial Reference System).*

## 8. Summary and Conclusion

The radiation level on a manned space flight to the Moon or Mars can vary from moderate over significant to deadly.

Moderate radiation levels can be expected when the Sun is almost calm. Then one may overcome a flight to the Moon and back with a moderate shielding without radiation damage. The shielding is only compulsory in the Van Allen radiation belt.

The flight path of Apollo 11 avoids the centre of the Van Allen radiation belt in an elegant way. It's a pity that this skilful trajectory has not been highlighted by NASA. For an even better avoidance one would have to fly first a high inclination orbit in order to then leave the Earth well above (or below) the Van Allen radiation belt; and finally to turn off in direction Moon – or Mars. But this would cost much more energy.

If the Sun suddenly got active, what cannot be predicted, also not for a short time span [lectures of solar researchers] & [13], one would rapidly be covered with a health affecting dose.

This substantial risk is confirmed by the following two statements of ESA [6] *„In the near-term, manned activities are limited to low altitude, and mainly low-inclination missions."* and [7] *"During the Apollo missions of the 1960s–70s, the astronauts were simply lucky not to have been in space during a major solar eruption that would have flooded their spacecraft with deadly radiation."* With other words the radiation risk of a manned lunar mission or beyond is regarded as not controllable.

In 1997 NASA proposed to increase the shielding by a factor of greater than 7 compared to the Apollo CM (as estimated in 1966 [18]): solar energetic particle events require a shielding of at least 52 mm aluminium or 100 mm water equivalent during transit to the Moon [19]. Then 2011 an ESA study concerning active radiation protection was made. The final documents describe principles based on huge superconducting magnets and they show a roadmap of more than 10 years up to first crewed flights. [20]

The radiation, specifically the massive rise from 500 to 1000 km altitude [Figures 4 and 5], is also a main reason why the International Space Station ISS remains between 300 and 400 km altitude.